\documentclass[acmsmall]{acmart}
\makeatletter
\let\@authorsaddresses\@empty
\makeatother
\renewcommand\footnotetextcopyrightpermission[1]{} 
\usepackage{graphicx}
\usepackage[caption=false,font=footnotesize,subrefformat=parens]{subfig}

\usepackage{amssymb}
\usepackage{xspace}
\usepackage{listings}
\usepackage{xcolor}

\newcommand{\fig}[1]{{Fig.~\ref{#1}}}
\newcommand{\eq}[1]{{Eqn.~\ref{#1}}}
\newcommand{\sect}[1]{{Sec.~\ref{#1}}}

\newcommand{\bm}[1]{{\boldsymbol{#1}}}
\def\eg{e.g.}
\def\ie{i.e.}
\def\etc{etc.}

\graphicspath{{./Figure/}}
\DeclareGraphicsExtensions{.pdf,.jpeg,.png}

\definecolor{codegreen}{rgb}{0,0.6,0}
\definecolor{codegray}{rgb}{0.5,0.5,0.5}
\definecolor{codepurple}{rgb}{0.58,0,0.82}
\definecolor{backcolour}{rgb}{0.95,0.95,0.93}
\ifodd 1
\newcommand{\com}[1]{\textbf{\color{red}(COMMENT: #1)}} 

\else

\newcommand{\com}[1]{}
\fi

\begin{document}

\title{Hands-on Wireless Sensing with Wi-Fi: A Tutorial}

\author{Zheng Yang}
\email{hmilyyz@gmail.com}
\affiliation{%
  \institution{Tsinghua University}
  \city{Beijing}
  \country{China}
}

\author{Yi Zhang}
\email{zhangyithss@gmail.com}
\affiliation{%
  \institution{Tsinghua University}
  \city{Beijing}
  \country{China}
}

\author{Guoxuan Chi}
\email{chiguoxuan@gmail.com}
\affiliation{%
  \institution{Tsinghua University}
  \city{Beijing}
  \country{China}
}

\author{Guidong Zhang}
\email{zhanggd18@gmail.com}
\affiliation{%
  \institution{Tsinghua University}
  \city{Beijing}
  \country{China}
}

\begin{abstract}
With the rapid development of wireless communication technology, wireless access points (AP) and internet of things (IoT) devices have been widely deployed in our surroundings. Various types of wireless signals (e.g., Wi-Fi, LoRa, LTE) are filling out our living and working spaces. 
Previous researches reveal the fact that radio waves are modulated by the spatial structure during the propagation process (e.g., reflection, diffraction, and scattering) and superimposed on the receiver. This observation allows us to reconstruct the surrounding environment based on received wireless signals, called ``wireless sensing''. 
Wireless sensing is an emerging technology that enables a wide range of applications, such as gesture recognition for human-computer interaction, vital signs monitoring for health care, and intrusion detection for security management.
Compared with other sensing paradigms, such as vision-based and IMU-based sensing, wireless sensing solutions have unique advantages such as high coverage, pervasiveness, low cost, and robustness under adverse light and texture scenarios. Besides, wireless sensing solutions are generally lightweight in terms of both computation overhead and device size.
This tutorial takes Wi-Fi sensing as an example. It introduces both the theoretical principles and the code implementation~\footnote{Code and data are available at \url{http://tns.thss.tsinghua.edu.cn/wst} and \url{http://tns.thss.tsinghua.edu.cn/widar3.0}.} of data collection, signal processing, features extraction, and model design. In addition, this tutorial highlights state-of-the-art deep learning models (e.g., CNN, RNN, and adversarial learning models) and their applications in wireless sensing systems.
We hope this tutorial will help people in other research fields to break into wireless sensing research and learn more about its theories, designs, and implementation skills, promoting prosperity in the wireless sensing research field.
\end{abstract}
\maketitle


\section{Wireless Sensing Background}
 
\subsection{What is Wireless Sensing}
 
Various sensors and sensor networks have thoroughly extended human perception of the physical world. Nowadays, numerous sensors have been deployed to complete various sensing tasks, resulting in a significant deployment and maintenance overhead. This problem becomes increasingly troublesome when a large sensing scale is in demand. Taking indoor person tracking as an example, a specialized tracking system only covers a room-level area, which is too small compared with the moving region during a person's daily life. Multiple tracking systems are needed to achieve practical sensing coverage in realistic living environments, \eg, houses, campuses, markets, airports, and offices, and the cost inevitably ramps up. 

Given the cost limitations of the sensors, many pioneers tried to figure out an alternative solution during the past decade. 
Nowadays, various types of signals (\eg, Wi-Fi, LoRa, LTE) are filling out our living and working spaces for wireless communication, which can be leveraged to capture the environmental changes without causing extra overhead. 
According to the electromagnetism theory, the radio signals emitted by the transmitter (Tx) experience various physical phenomena such as reflection, diffraction, and scattering during the propagation process and form into multiple propagation paths.
In this way, the superimposed multipath signals collected by the receiver carry spatial information about the signal propagation environment. Relying only on the ambient wireless signals and ubiquitous communication devices, wireless sensing emerges as a novel paradigm for environment sensing. 


In recent years, wireless sensing technology has attracted many research interests to bring wireless sensing from the imagination into reality, by boosting sensing granularity, improving system robustness, and exploring application scenarios. Many of their works on wireless sensing have been published in flagship conferences and journals, such as ACM SIGCOMM, ACM MobiCom, ACM MobiSys, IEEE INFOCOM, USENIX NSDI, IEEE/ACM ToN, IEEE JSAC, and IEEE TMC. 
In addition, many famous companies are also exploring the productization of sensorless sensing, launching various IoT devices for human-computer interaction,  security monitoring, and health care.

\subsection{Comparison of Wireless Sensing and Computer Vision}
 
Typical RF signals (300 kHz - 300 GHz) and visible light signals (380 THz - 750 THz) are essentially electromagnetic (EM) waves. When propagating in our physical world, the EM waves experience a variety of physical phenomena such as reflection, diffraction, and scattering. Multipath signals are eventually superimposed and received by the receiver. Therefore, the received superimposed signals carry the physical information of the signal propagation space.

Both the RF-based and the vision-based sensing algorithms share similar processes. They first analyze the received signals (radio signal at the antenna or visible light at the camera lens), from which the features reflecting the propagation space are extracted and finally resolved by algorithms to realize the sensing of the surrounding environment.

Compared with vision-based sensing, wireless sensing solutions have unique advantages such as high coverage~\cite{chi2021locate}, pervasiveness, low cost, and robustness under adverse light and texture scenarios~\cite{chi2022widrone}.

\begin{figure}[t]
\centering
\includegraphics[width=.9\columnwidth]{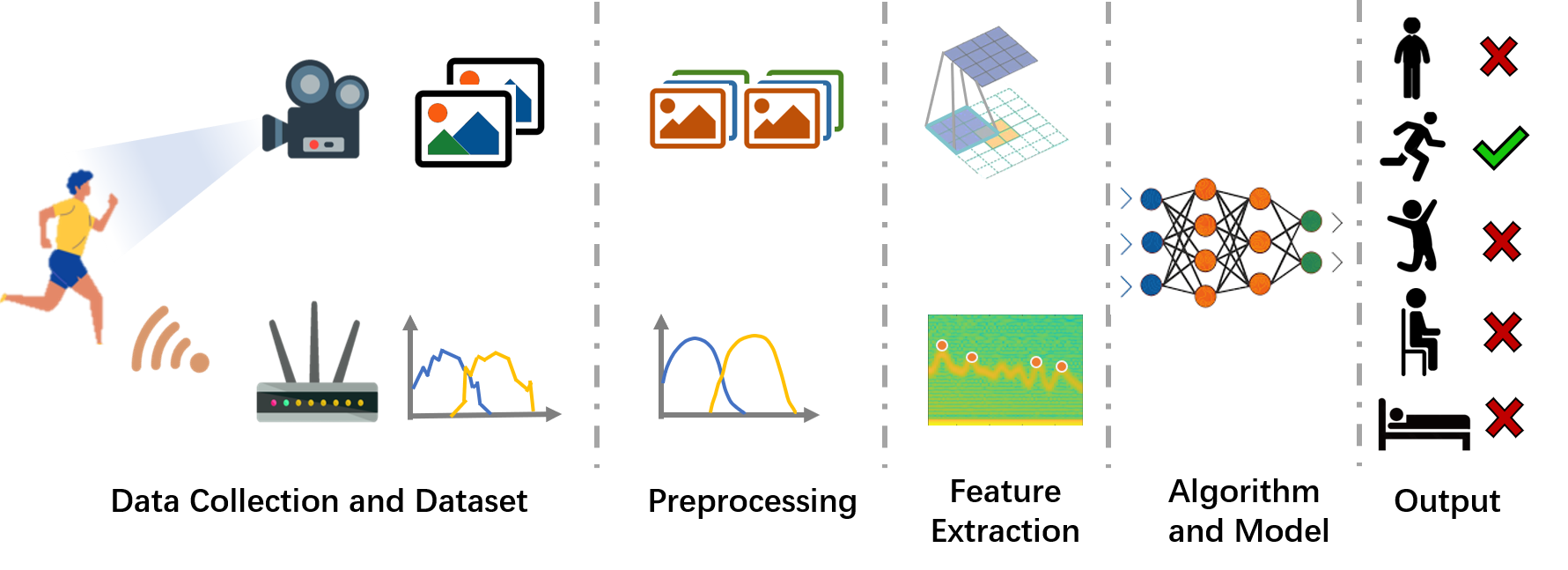}
\caption{Comparision of vision-based sensing and RF-based sensing processes.}
\label{fig:sensing-proc}
\end{figure}

 \subsection{Wireless Sensing Applications}
 
Wireless sensing systems are capable of perceiving changes in surrounding environments, objects, and human bodies. In this subsection, we take \textit{passive human sensing} applications as an example, which refers to a human-centered sensing application that doesn't require the user to carry any device. Therefore, such an sensing application is also termed as \textit{device-free sensing} or \textit{non-invasive sensing}. Passive human sensing enables a wide range of applications, including smart homes, security surveillance, and health care.

In \textbf{smart home} applications, passive human sensing recognizes a person's behavior or intention based on the user's physical locations, gestures, and postures. Passive human sensing brings a better user experience without imposing restrictions on the user. For example, users can remotely control electrical devices, \eg, television, computer, or washer, by merely performing gestures in the air~\cite{Widar3.0, abdelnasser2015wigest}. Likewise, when playing video games, users can interact with the computer by performing different postures~\cite{WiPose}.

In \textbf{security surveillance} applications, traditional methods adopt infrared or RGB cameras to monitor illegal invasions, protect valuable properties, and deal with emergencies. However, cameras are constrained by the limited field of view or blockage of opaque or metallic objects, rendering these methods to fail when the target is not in the Line-of-Sight (LoS) area of the surveillance camera or hidden behind other objects. In contrast to visual surveillance, wireless sensing technology leverages radio signals, which provide omnidirectional coverage around the wireless devices and are less prone to blockages. For example, wireless signals can be used to detect illegal intrusions~\cite{deman2015jsac,pads2014icpads}.
Besides, they can also be used to detect if properties have been moved from their original places. 

In \textbf{health care} applications, passive human sensing can be leveraged to detect vital signals such as human respiration, heartbeat, gait, and accidental fall. Specifically, some researchers have exploited Wi-Fi signals to detect human respiration~\cite{respiration2016daqing} for sleep monitoring. Some other works~\cite{zhang2020gaitid, gaitway} have extracted gait patterns from Wi-Fi signals to recognize human identity. Recently, Wi-Fi signals have been further used to detect accidental falls to relieve the need for wearable sensors~\cite{defall, falldefi}.
\section{Understanding CSI}

Channel state information (CSI) lays the foundation of most wireless sensing techniques, including Wi-Fi sensing, LTE sensing, and so on. CSI provides physical channel measurements in subcarrier-level granularity, and it can be easily accessed from the commodity Wi-Fi network interface controller (NIC). 

CSI describes the propagation process of the wireless signal and therefore contains geometric information of the propagation space. Thus, understanding the mapping relationship between CSI and spatial geometric parameters lays the foundation for feature extraction and sensing algorithm design.

This section focuses on two mainstream CSI models: the ray-tracing model and the scattering model. The two models are based on two perspectives of understanding the signal propagation process. Thus, they have unique advantages and apply to different scenarios.

\subsection{Ray-tracing Model}

In typical indoor environments, a signal sent by the transmitter arrives at the receiver via multiple paths due to the reflection of the radio wave. Along each path, the signal experiences a certain attenuation and phase shift. The received signal is the superimposition of multiple alias versions of the transmitted signal. Therefore, the complex baseband signal strength measured at the receiver at a specific time can be written as follows~\cite{yang2013rssi}:
\begin{equation}
    V=\sum_{n=1}^{N}{\lVert V_n \rVert e^{-j\phi_n}},
\label{eq:p2c2-1}
\end{equation}
where $V_n$ and $\phi_n$ are the amplitude and phase of the $n^{th}$ multipath component (note that the modulation scheme of the signal is implicitly considered), and $N$ is the total number of components. On this basis, the recieve signal strength indicator (RSSI) can be written as the received power in decibels (dB):
\begin{equation}
    \mathrm{RSSI} = 10\log_2\left( \lVert V \rVert^2 \right).
\label{eq:p2c2-2}
\end{equation}

As the superimposition of multipath components, RSSI not only varies rapidly with propagation distance changing at the order of the signal wavelength but also fluctuates over time, even for a static link. A slight change in specific multipath components may result in significant constructive or destructive multipath components, leading to considerable fluctuations in RSSI.

The essential drawback of RSSI is the failure to reflect the multipath effect. The wireless channel is modeled as a linear temporal filter to fully characterize individual paths, known as channel impulse response (CIR). Under the time-invariant assumption, CIR $h(t)$ is represented as:
\begin{equation}
    h(t) = \sum_{n=1}^{N}{\alpha_n e^{-j\phi_n} \delta(t - \tau_n)},
\label{eq:p2c2-3}
\end{equation}
where $\alpha_n$, $\phi_n$, and $\tau_n$ are the complex antenuation, phase, and time delay of the $n^{th}$ path, respectively. $N$ is the total number of multipath and $\delta(\cdot)$ is the Dirac delta function. Each impulse represents a delayed multipath component, multiplied by the corresponding amplitude and phase.

\begin{figure}[t]
\centering
\includegraphics[scale=.98]{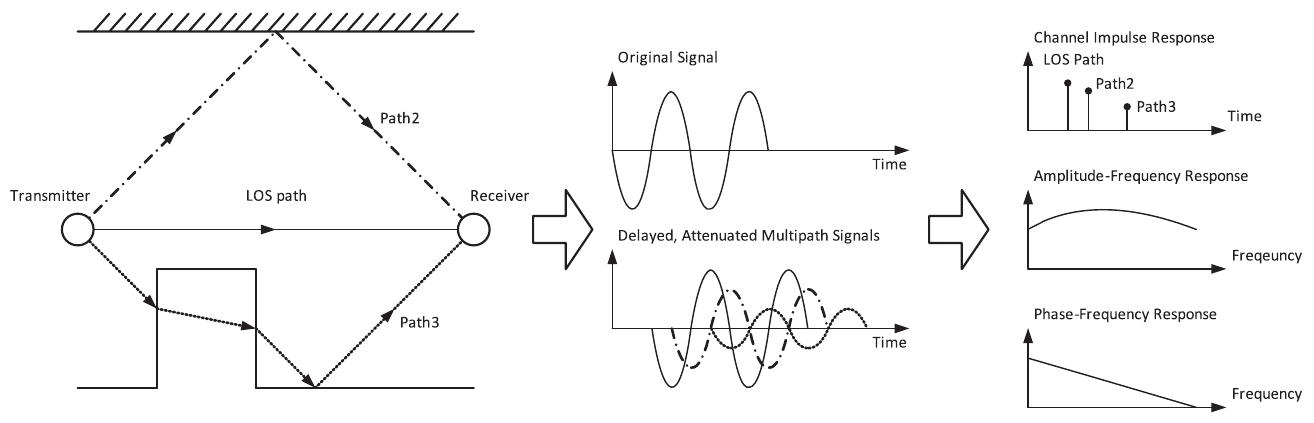}
\caption{Multipath propagations, received signals, and channel responses.}
\label{fig:multipath_model}
\end{figure}

In the frequency domain, the multipath causes frequency-selective fading, which is characterized by channel frequency response (CFR). CFR is essentially the Fourier transform of CIR. It consists of both the amplitude response and the phase response. \fig{fig:multipath_model} demonstrate a multipath scenario, the transmitted signal, the received signal, and the illustrative channel responses. Both CIR and CFR depict a small-scale multipath effect and are used for fine-grained channel measurement. Note that the complex amplitudes and antennuation are concerned in CIR and CFR, while another pair of parameters in terms of the signal power is Power Delay Profile (PDP) and Power Spectrum Density (PSD).

CIR and CFR are measured by decoupling the transmitted signal from the received signal. Specifically, in the time domain, the received signal $r(t)$ is the convolution of transmitted signal $s(t)$ and channel impulse response $h(t)$:
\begin{equation}
    r(t) = s(t) \otimes h(t),
\label{eq:p2c2-4}
\end{equation}
which indicates the recieved signal is generated from the transmit signal after it propagating from multipath channel.

Similarly, in the frequency domain, the received signal spectrum $R(f)$ is the multiplication of the transmitted signal spectrum $S(f)$ and the channel frequency response $H(f)$:
\begin{equation}
    R(f) = S(f) H(f).
\label{eq:p2c2-5}
\end{equation}
Note that the $R(f)$ and $S(f)$ are the Fourier transform of the recieved signal $r(t)$ and the transmitted signal $s(t)$ respectively. In this way, \eq{eq:p2c2-4} and \eq{eq:p2c2-5} forms a beautiful ``symmetric'' relationship.

As demonstrated in \eq{eq:p2c2-4} and \eq{eq:p2c2-5}, CIR can be derived from the deconvolution operation of received and transmitted signals, and CFR can be treated as the ratio of the received and the transmitted spectrums. Compared with multiplication, the convolution operation is generally time-consuming. Therefore, in most cases, the device focuses on calculating the CFR, and the CIR can be further derived from the CFR using the inverse Fourier transform~\cite{10.1145/1287853.1287867}:
\begin{equation}
    h(t) = \frac{1}{P_s}\mathfrak{F}^{-1}\left\{ S^*(f)R(f) \right\},
\label{eq:p2c2-6}
\end{equation}
where $\mathfrak{F}^{-1}$ denotes the inverse Fourier transform, $*$ is the conjugate operator, and $P_s$ approximates the transmitted signal power.

Although the derivation of CIR and CFR is independent of the modulation scheme, it could be more convenient to implement the process on commercial devices with specific modulation schemes. For the wireless standard where the orthogonal frequency division modulation (OFDM) is adopted (\eg, 802.11a/g/n/ac/ax), the amplitude and phase sampled on each subcarrier can be treated as a sampled version of the signal spectrum $S(f)$. On this basis, a sampled version of $H(f)$ can be easily get from the OFDM receivers.

Recent advances in the wireless community make it possible to get the sampled version of CFR from commercial-off-the-shelf (COTS) Wi-Fi NICs. 
The extracted CFR are often refered to as a series of complex numbers, which depicts the amplitude and phase of each subcarrier:
\begin{equation}
    H(f_j)=||H(f_j)||e^{\angle H(f_j)},
\label{eq:p2c2-7}
\end{equation}
where $H(f_j)$ is a sample at the $j^{th}$ subcarrier, with $\angle H(f_k)$ denotes its phase.
Most research paper treat the CFR at sampled at different subcarriers as the CSI data, which can be written as $\bm{H} = \left\{ H(f_j) \ \vert \ j \in [1, J], j \in \mathbb{N} \right\}$, where $J$ denotes the total number of subcarriers.

The ray-tracing model establishes the relationship between geometric properties of the signal propagation space and the CSI data. Theoretically, by analyzing the multipath signal, various types of geometric information (\eg, the propagation distance, the reflection points) can be derived. Therefore, the ray-tracing model is widely adopted in wireless localization and tracking tasks. In addition, many wireless detection and sensing systems also extract geometric features based on the ray-tracing models, and further put them into machine learning models for regression or classification.

A significant drawback of the ray-tracing model is that it is based on a simple environmental assumption. Therefore, in a complex environment where the signal undergoes diffraction or is disturbed by various types of noise, it becomes difficult to accurately recover all the spatial information by only limited CSI data.

\subsection{Scattering Model}

The above-mentioned ray-tracing model characterizes signals with multiple propagation paths, which may not apply to rich-scattering environments.
To enable wireless sensing in more complex scenarios, some works~\cite{wispeed, gaitway} establish the scattering model for CSI measurements with Wi-Fi.

The CSI data reported by commodity Wi-Fi NIC can be modeled as:
\begin{equation}
H(f,t) = \sum_{n=1}^{N}{\alpha_n(f,t)e^{-j\phi_n(f, t)}},
\label{eq:csi}
\end{equation}
where $N$ is the total number of reflection paths, $\alpha_n$ is the amplitude attenuation of path $n$, $\phi_n$ is the corresponding phase.

\begin{figure}[tb]
\centering
\includegraphics[width=0.6\columnwidth]{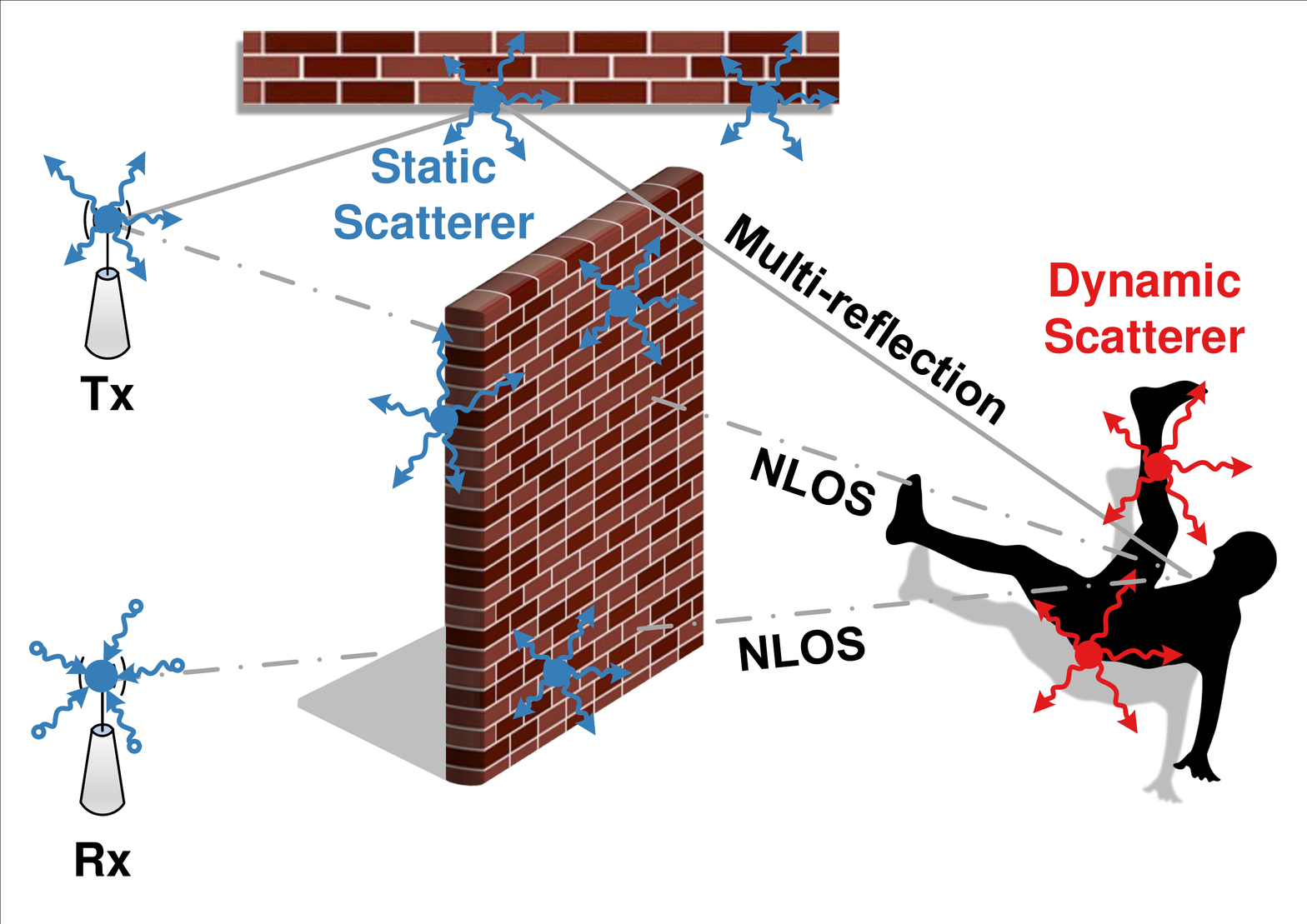}
\caption{Wi-Fi signals in rich-scattering environment}
\label{fig:scatterer}
\end{figure}

As shown in \fig{fig:scatterer}, the scattering model treats all the objects in indoor environments as \textit{scatterers} that diffuse the signals to all directions. The CSI observed by the receiver is added up with the portions contributed by the static (furniture, walls, \etc) and dynamic (arms, legs, \etc) scatterers. Intuitively, each scatter is deemed as a virtual Tx. Such modeling can be applied to typical indoor scenarios, where the rooms are crowded with furniture, and signals could propagate in almost all directions. On this basis, we can dismiss the specific signal propagation path and only statistically investigate the relationship between the observed CSI and the moving speed. 

We decompose the observed CSI into the portions contributed by individual scatterers:
\begin{equation}
\label{eq:csi_scatterers}
H(f,t) = \sum_{o \in \Omega_s(t)}{H_o(f,t)} + \sum_{p \in \Omega_d(t)}{H_p(f,t)},
\end{equation}
where $\Omega_s(t)$ and $\Omega_d(t)$ are the sets of static and dynamic scatterers, $H_p(f,t)$ is the portion of observed CSI contributed by $p^{th}$ dynamic scatterer. For each scatterer, the diffused components undergo anonymous propagation processes and eventually add up at the receiver. Try to a three-dimensional coordinate with its origin at the $p^{th}$ scatterer and its z-axis align with the scatterer's moving direction, the representation of $H_p(f,t)$ is further decomposed as follows~\cite{wispeed}:
\begin{equation}
\label{eq:csi_angular}
H_p(f,t) = \int_{0}^{2\pi}\int_{0}^{\pi}{h_p(\alpha,\beta,f,t)\exp(-jkv_p\cos(\alpha)t)\mathrm{d} \alpha \mathrm{d} \beta},
\end{equation}
where $k=2\pi/\lambda$ and $\lambda$ is the signal wavelength, $v_p$ is the speed of the $p^{th}$ dynamic scatterer, $\alpha$ and $\beta$ are the azimuth and elevation angles, $\exp(-jkv_i\cos(\alpha)t)$ represents the phase shift of the signal on direction $(\alpha,\beta)$. $h_p(\alpha,\beta,f,t)$ is the portion of observed CSI contributed by $p^{th}$ scatterer on direction $(\alpha,\beta)$. Inherited from the physical properties of EM waves~\cite{EMincavities}, $h_p(\alpha,\beta,f,t)$ can be viewed as a circularly-symmetric Gaussian random variable. For $\forall p, q \in \Omega_d$ and $(\alpha_1,\beta_1) \neq (\alpha_2,\beta_2)$, $h_p(\alpha_1,\beta_1,f,t)$ is independent of $h_q(\alpha_2,\beta_2,f,t)$.

Based on \eq{eq:csi_scatterers} and \eq{eq:csi_angular}, we are able to derive the statistical property by analyzing the autocorrelation function (ACF) of $H(f,t)$~\cite{EMincavities}:
\begin{equation}
\label{eq:csi_acf}
\begin{split}
\rho_{H}(f,\tau)&\triangleq \frac{\mathrm{Cov}[H(f,t),H(f,t+\tau)]}{\mathrm{Cov}[H(f,t),H(f,t)]}\\
&=\frac{\sum_{p \in \Omega_d}{\sigma_p^2(f) \mathrm{sinc}(kv \tau)}}{\sum_{p \in \Omega_d}{\sigma_p^2(f)}}
\approx \mathrm{sinc}(kv\tau),
\end{split}
\end{equation}
where $\mathrm{Cov}[\cdot,\cdot]$ is the covariance between two random variables, $\sigma_p^2(f)$ is the variance of $h_p(\alpha,\beta,f,t)$, $\mathrm{sinc}(kv \tau) = \frac{\sin(kv \tau)}{kv \tau}$, and $\tau$ is the time lag. 

For some human activities like fall and walking, we can assume that the dynamic scatterers are mainly on the torso and have similar speeds $v$.
Thus, \eq{eq:csi_acf} is approximated with a much simpler form: $\rho_{H}(f,\tau) \approx \mathrm{sinc}(kv\tau)$. Using this formulation, we have quantitatively established the relationship between the huamn speed and the ACF of CSI. In practice, the speed $v$ is extracted by matching the first peak of the $\mathrm{sinc}(x)$ function and the first peak of ACF:
\begin{equation}
\label{eq:csi_peak_matching}
v=\frac{x_0}{k \tau_0}=\frac{x_0 \lambda}{2\pi \tau_0},
\end{equation}
where $x_0$ is the constant value representing the location of the first peak of $\mathrm{sinc}(x)$ function, $\tau_0$ is the time lag corresponding to the first peak of the ACF.

\begin{figure}[tb]
\centering
\includegraphics[width=0.6\columnwidth]{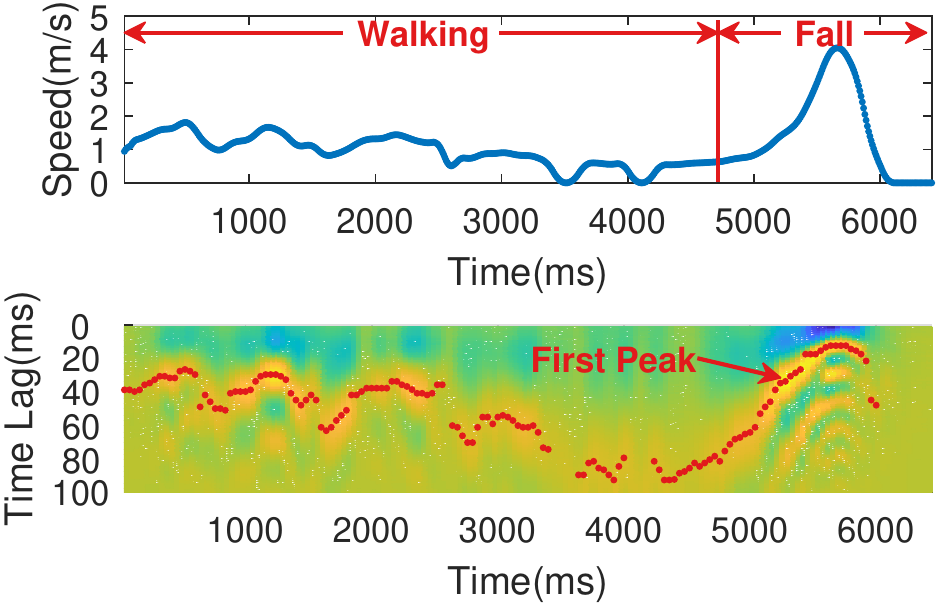}
\caption{The estimated speed and the ACF of CSI signals}
\label{fig:wispeed_SM}
\end{figure}

As a straightforward example, we let a volunteer to walk and fall in a heavily furnished room and extract the speed with the CSI collected from a Wi-Fi link located in another room. As is shown in \fig{fig:wispeed_SM}, even with non-line-of-sight (NLoS) occlusions and multipath pollution, the walking speed and fall speed are estimated consistently, demonstrating its robustness to environmental diversity.

The scattering model is generally applicable to various speed-oriented sensing The scattering model generally applies to various speed-oriented sensing tasks such as intrusion detection and fall detection. As a statistical model, it has unique advantages in complex scenarios.

However, the scattering model fails to construct a correspondence between CSI and geometric parameters directly, and thus is generally not applicable to precise localization and tracking tasks.

\section{CSI Data Collection}

CSI data collection is the first step toward the implementation of a practical wireless sensing system. This section introduces one of the most famous Wi-Fi sensing datasets, the Widar3 dataset, so that beginners can quickly get start with wireless sensing with minimal effort.
This section also introduces three of the most famous CSI collection tools that researchers can use to try out collecting CSI data with COTS NICs.

\subsection{Widar3 Dataset}

Open datasets are essential to provide comprehensive knowledge for model training and a unified benchmark for model comparison. Open datasets are even more necessary in the wireless sensing field because RF signals are more sensitive to devices and deployment environments. However, the absence of high-quality and large-scale datasets has become the bottleneck that hindered the progress of wireless sensing technology. Existing wireless sensing datasets suffer from small scales and limited scenarios in 2019 when we started to build the Widar3 dataset. Widar3\footnote{\url{http://tns.thss.tsinghua.edu.cn/widar3.0}} is a wireless sensing dataset for human activity recognition. It is collected from commodity Wi-Fi NICs in the form of RSSI and CSI. It consists of 258,000 instances of hand gestures with a duration of 8,620 minutes and from 75 domains. Widar3 is so far the largest and most comprehensive dataset in this field and receives widespread attention from researchers all over the world. Widar3 dataset is publicly available at IEEE DataPort (official data repository) and continues evolving to contain more types of activities.

\subsection{PicoScenes Platform}
PicoScenes~\cite{jiang2021eliminating} is a versatile and powerful middleware for CSI-based Wi-Fi sensing research. 
It is one of the very few tools that support the latest 802.11ac/ax protocols.
It supports many prevalent commercial NICs, including Qualcomm Atheros AR9300 (QCA9300), Intel Wireless Link 5300 (IWL5300), Intel AX200 and Intel AX210.
PicoScenes supports up to 27 NICs to work concurrently for packet injection and CSI measurement.

PicoScenes is architecturally versatile and flexible. It encapsulates all the low-level features into unified and hardware-independent APIs and exposes them to the upper-level plugin layer. As a result, users can quickly prototype their own measurement plugins.

The data reported by PicoScenes can be parsed in MATLAB as a struct, containing the CSI data of different packets, subcarriers, and antennas. The struct also includes other helpful information such as timestamps, RSSI, and the signal-to-noise ratio (SNR).

The homepage\footnote{\url{https://ps.zpj.io}} of this tool can be accessed for more detailed information.

\subsection{Intel 5300 NIC CSI Tool}

This CSI Tool~\cite{halperin2011tool} is built upon the Intel WiFi Wireless Link 5300 802.11n MIMO radios, using a modified firmware and the open-source Linux wireless driver. It includes all the software and scripts required to collect, read, and parse CSI.

The IWL5300 provides 802.11n CSI of 30 subcarrier groups. Each group contains 2 adjacent subcarriers given 20 MHz bandwidth or 4 given 40 MHz bandwidth. Each CSI sample is a complex number, with a signed 8-bit resolution for both real and imaginary parts. One CSI record is a $A \times 30$ matrix, where $M$ is the number of pairs of transmitting and receiving antennas. 

The homepage\footnote{\url{https://dhalperi.github.io/linux-80211n-csitool}} of this tool can be accessed for detailed information.

\subsection{Atheros CSI Tool}

Atheros CSI Tool~\cite{xie2018precise} is an open-source 802.11n measurement and experimentation tool. It enables the extraction of detailed PHY wireless communication information from the Atheros WiFi NICs, including the Channel State Information (CSI), the received packet payload, and other information (the time stamp, the RSSI of each antenna, the data rate, \etc). Atheros-CSI-Tool is built on top of ath9k, an open-source Linux kernel driver supporting Atheros 802.11n PCI/PCI-E chips. Thus, this tool theoretically supports all types of Atheros 802.11n WiFi chipsets. We have tested it on Atheros AR9580, AR9590, AR9344, and QCA9558. Furthermore, Atheros CSI Tool is open source, and all functionalities are implemented in software without any modification to the firmware. Therefore, one can extend the functionalities of Atheros CSI Tool with their own codes under the GPL license. 

Atheros-CSI-Tool works on various Linux distributions, \eg, Ubuntu, OpenWRT, Linino, \etc. Different Linux distribution works with different hardware. Ubuntu works for personal computers like laptops or desktops. OpenWRT works for embedded devices such as WiFi routers. Linino works for IoT devices, such as Arduino YUN. The official website provides the source code for the Ubuntu version and OpenWRT version of the Atheros CSI tool.

The homepage\footnote{\url{https://wands.sg/research/WiFi/AtherosCSI}} of this tool can be accessed for detailed information.
\section{CSI Feature Extraction}

The CSI features lay the fundation of wireless sensing. In particular, for different sensing tasks, choosing the most appropriate features can effectively improve the system performance.
In addition, the quality of the extracted features determines the effectiveness of the sensing system. 

For ease of illustration, code implementation used for feature extraction is provided below, which is a \lstinline{main} function to call different function in the following subsections.

\begin{lstlisting}[language=MATLAB]
%{
  CSI Feature Extraction for Wi-Fi sensing.  
  - Input: csi data used for calibration, and csi data that need to be sanitized.
  - Output: sanitized csi data.

  To use this script, you need to:
  1. Make sure the csi data have been saved as .mat files.
  2. Check the .mat file to make sure they are in the correct form. 
  3. Set the parameters.
  
  Note that in this algorithm, the csi data should be a 4-D tensor with the size of [T S A L]:
  - T indicates the number of csi packets;
  - S indicates the number of subcarriers;
  - A indicates the number of antennas (i.e., the STS number in a MIMO system);
  - L indicates the number of HT-LTFs in a single PPDU;
  Say if we collect a 10 seconds csi data steam at 1 kHz sample rate (T = 10 * 1000), from a 3-antenna AP (A = 3),  with 802.11n standard (S = 57 subcarrier), without any extra spatial sounding (L = 1), the data size should be [10000 57 3 1].
%}

clear all;
addpath(genpath(pwd));

%% 0. Set parameters.
% Path of the raw CSI data.
src_file_name = './data/csi_src_test.mat';

% Speed of light.
global c;
c = physconst('LightSpeed');
% Bandwidth.
global bw;
bw = 20e6;
% Subcarrier frequency.
global subcarrier_freq;
subcarrier_freq = linspace(5.8153e9, 5.8347e9, 57);
% Subcarrier wavelength.
global subcarrier_lambda;
subcarrier_lambda = c ./ subcarrier_freq;

% Antenna arrangement.
antenna_loc = [0, 0, 0; 0.0514665, 0, 0; 0, 0.0514665, 0]';
% Set the linear range of the CSI phase, which varies with NIC types.
linear_interval = (20:38)';

%% 1. Read the csi data for calibration and sanitization.
% Load the raw CSI data.
csi_src = load(src_file_name).csi;      % Raw CSI.

%% 2. Perform various wireless sensing tasks.
% Test example 1: angle/direction estimation with imperfect CSI.
[packet_num, subcarrier_num, antenna_num, ~] = size(csi_src);
aoa_mat = naive_aoa(csi_src, antenna_loc, zeros(3, 1));
aoa_gt = [0; 0; 1];
error = mean(acos(aoa_gt' * aoa_mat));
disp("Angle estimation error: " + num2str(error));

% Test example 2: distance estimation with CSI.
tof_mat = naive_tof(csi_src);
est_dist = mean(tof_mat * c, 'all');
disp("The ground truth distance is: 10 m");
disp("The estimated distance is: " + num2str(est_dist) + " m");
\end{lstlisting}

\subsection{Time of Flight}

\begin{figure}[h]
\centering
\includegraphics[width=.95\textwidth]{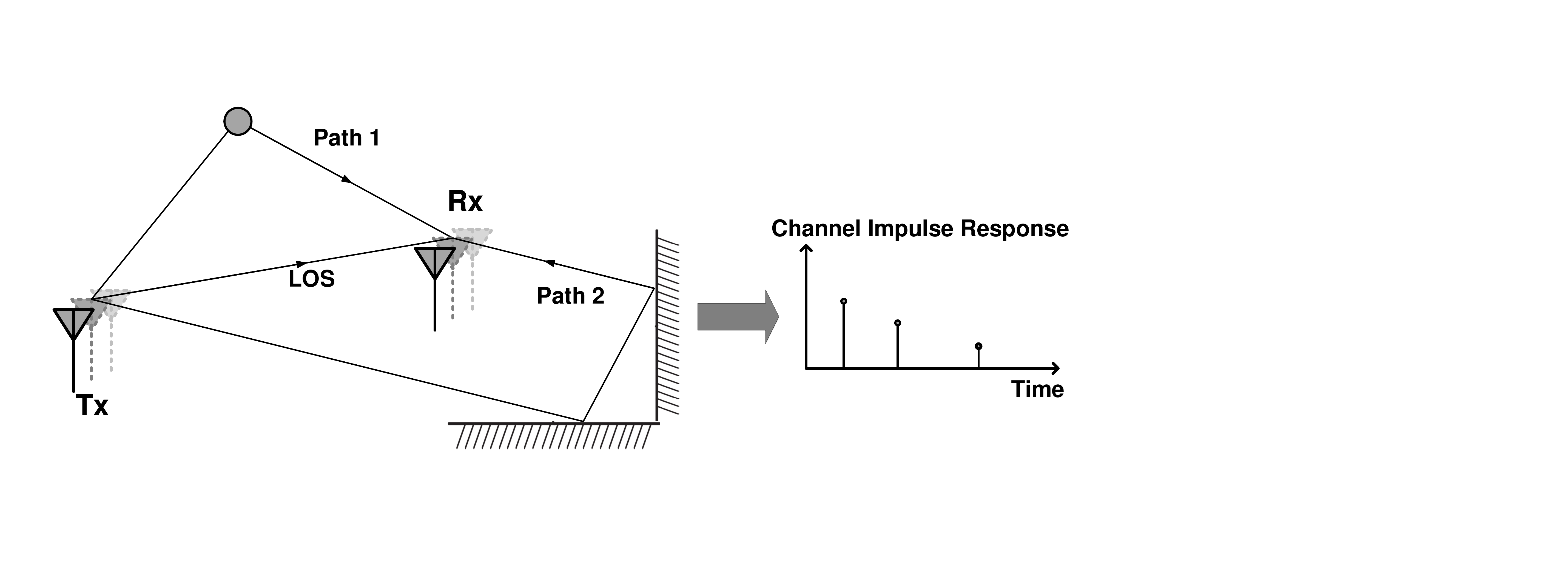}
\caption{The relationship between ToF and CIR.}
\label{fig:p2c2-tof}
\end{figure}

ToF is the time duration the signal propagates from the transmitter to the receiver along a specific path.
Given the frequency $f$, the phase shift introduced by the ToF $\tau$ is:
\begin{equation}
\phi_{\mathrm{ToF}} = -2\pi f\tau
\end{equation}
As the superimposition of multipath signals, CSI can be represented based on the ray-tracing model:
\begin{equation}
    H(f)=\sum_{n=1}^{N}\alpha_{n}e^{-j2\pi f\tau_n}
    \label{eq:p2c2_tof}
\end{equation}
where $N$ is the total number of multipath, and $\alpha_n$ and $\tau_n$ are the complex attenuation factor and time of flight (ToF) for the $n^{th}$ path, respectively.
Theoretically, the ToF of all paths can be identified in CIR, which can be calculated by applying the inverse Fourier transform to CSI samples of all subcarriers. 
However, since the transmitter and the receiver lack synchronization, non-zero temporal shifts exist in CIR, and the absolute ToF is typically not accurate enough. The limited bandwidth also constrains the time resolution, causing meter-level ToF ambiguity 
The relationship between signal propagation path, ToF, and CIR is shown in Figure~\ref{fig:p2c2-tof}.

The following function \lstinline{naive_tof} intends to extract the ToF of the strongest path (typically the shortest path) based on inverse Fourier transform.

\begin{lstlisting}[language=MATLAB]
function [tof_mat] = naive_tof(csi_data)
    % naive_tof
    % Input:
    %   - csi_data is the CSI used for ranging; [T S A E]
    %   - ifft_point and bw are the IFFT and bandwidth parameters;
    % Output:
    %   - tof_mat is the rough time-of-flight estimation result; [T A]

    global c, bw;
    [pakcet_num, subcarrier_num, antenna_num, extra_num] = size(csi_data);
    ifft_point = power(2, ceil(log2(subcarrier_num)));
    % Get CIR from each packet and each antenna by ifft(CFR);
    cir_sequence = zeros(packet_num, antenna_num, extra_num, ifft_point);

    for p = 1:packet_num
        for a = 1:antenna_num
            for e = 1:extra_num
                cir_sequence(p, a, e, :) = ifft(csi_data(p, :, a, e), ifft_point);
            end
        end
    end
    cir_sequence = squeeze(mean(cir_sequence, 4)); % [T ifft_point A]
    half_point = ifft_point / 2;
    half_sequence = cir_sequence(:, 1:half_point, :); % [T half_point A]
    peak_indices = zeros(packet_num, antenna_num); % [T A]
    for p = 1:pakcet_num
        for a = 1:antenna_num
            [~, peak_indices(p, a)] = max(half_sequence, [], 2);
        end
    end
    % Calculate ToF;
    tof_mat = peak_indices .* subcarrier_num ./ (ifft_point .* bw); % [T A]
end
\end{lstlisting}

\subsection{Angle of Arrival and Angle of Departure}

\begin{figure}[h]
\centering
\includegraphics[width=.65\textwidth]{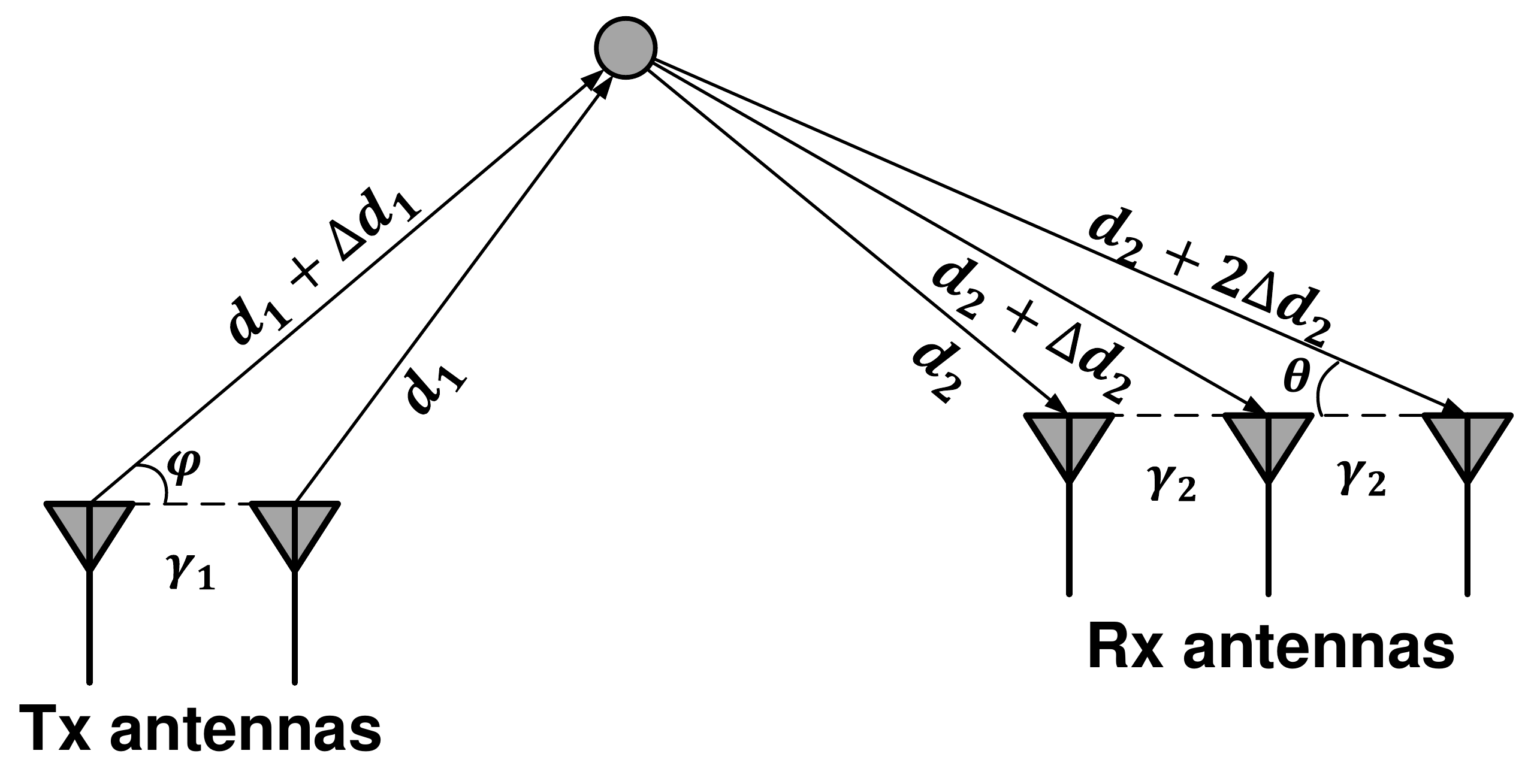}
\caption{Angle of Arrival and Angle of Departure.}
\label{fig:p2c2-aoa}
\end{figure}

When a NIC equipes with multiple antennas, a local coordinate at the device can be created. 
As shown in Figure~\ref{fig:p2c2-aoa}, for a transmitter, the angle of departure (AoD) $\varphi$ represents the direction in the local coordinate along which the transmitted signal is emitted. 
For a receiver, the angle of arrival (AoA) $\theta$ represents the direction in the local coordinate along which the received signal is captured. 
Since the antennas are spatially separated, non-zero phase shifts between antennas are introduced. The phase shifts depend on the AoA/AoD.
Specifically, suppose the relative location between two antennas is $\bm{\Delta{l}}=(\Delta_x, \Delta_y)$ and the unit direciton vector of AoA is $\bm{e}=(cos\theta, \sin\theta)$, the phase shift between the two antennas is: 
\begin{equation}
\label{eqn:naive-aoa}
\phi_{\mathrm{AoA}} = \frac{2\pi}{\lambda}\bm{\Delta{l}}\cdot\bm{e}
\end{equation}
Then CSI can be modeled as:
\begin{equation}
H(k)=\sum_{n=1}^{N}\alpha_{n}e^{-j\frac{2\pi}{\lambda}\bm{\Delta{l}}\cdot\bm{e}}
\end{equation}
where $k$ represents the $k^{th}$ antenna at the receiver. The same model applies to the AoD at the trasmitter side and the 3D space with azimuth and elevation angles.

In practice, algorithms such as Capon~\cite{CAPON} and MUSIC~\cite{MUSIC} can be used to estimate the AoA/AoD of multiple paths from the CSI of the antenna array.

MUSIC analyses the incident signals on multiple antennas to find out the AoA of each signal.
Specifically, suppose $D$ signals $F_{1},\cdots,F_{D}$ arrive from directions $\theta_{1},\cdots,\theta_{D}$ at $M > D$ antennas.
The received signal at the $k^{th}$ antenna element, denoted as $X_{k}$, is a linear combination of the $D$ incident wavefronts and noise $W_{k}$:
\begin{equation*}
\left[\begin{array}{c}
X_{1}\\
X_{2}\\
\vdots\\
X_{M}
\end{array}\right]
=
\left[\begin{array}{llll}
\bm{a}(\theta_{1})&\bm{a}(\theta_{2})&\ldots&\bm{a}(\theta_{D})
\end{array}\right]
\left[\begin{array}{c}
F_{1}\\
F_{2}\\
\vdots\\
F_{D}
\end{array}\right]
+
\left[\begin{array}{c}
W_{1}\\
W_{2}\\
\vdots\\
W_{M}
\end{array}
\right]
\end{equation*}
or
\begin{equation}
\bm{X}=\bm{AF}+\bm{W}
\end{equation}
where $\mathbf{a}(\theta_{k})$ is the array steering vector that characterizes added phase (relative to the first antenna) of each receiving component at the $k^{th}$ antenna.
$\bm{A}$ is the matrix of steer vectors.
As shown in Figure~\ref{fig:p2c2-aoa}, for a linear antenna array with elements well synchronized,
\begin{equation}\label{eq:aoa_phase}
\bm{a}(\theta)=\left[
\begin{array}{c}
1\\
e^{-j\frac{2\pi}{\lambda}\bm{\Delta{l(1)}}\cdot\bm{e}}\\
e^{-j\frac{2\pi}{\lambda}\bm{\Delta{l(2)}}\cdot\bm{e}}\\
\vdots\\
e^{-j\frac{2\pi}{\lambda}\bm{\Delta{l(M-1)}}\cdot\bm{e}}
\end{array}
\right]
\end{equation}
Suppose $W_{k}\sim N(0, \sigma^{2})$, and $F_k$ is a wide-sense stationary process with zero mean value, the $M\times M$ covariance matrix of the received signal vector $\bm{X}$ is:
\begin{equation}
\begin{aligned}
\bm{S}&=\overline{\bm{XX}^{\mathrm{H}}}\\
&=\bm{A}\overline{\bm{FF}^{\mathrm{H}}}\bm{A}^{\mathrm{H}}+\overline{\bm{WW}^{\mathrm{H}}}\\
&=\bm{APA}^{\mathrm{H}} + \sigma^{2}\bm{I}
\end{aligned}
\end{equation}
where $\bm{P}$ is the covariance matrix of transmission vector $\bm{F}$. The notation $(\cdot)^{\mathrm{H}}$ represents conjugate transpose and $\overline{(\cdot)}$ represents expectation.

The covariance matrix $\bm{S}$ has $M$ eigenvalues $\lambda_{1},\cdots,\lambda_{M}$ associated with $M$ eigenvectors $\bm{e}_{1},\bm{e}_{2},\cdots,\bm{e}_{M}$.
Sorted in a non-descending order, the smallest $M-D$ eigenvalues correspond to the noise while the rest $D$ correspond to the $D$ incident signals.
In other word, the $M$-dimension space can be divided into two orthogonal subspace, the noise subspace $\bm{E}_{N}$ expanded by eigenvectors $\bm{e}_{1},\cdots,\bm{e}_{M-D}$, and the signal subspace $\bm{E}_{S}$ expanded by eigenvectors $\bm{e}_{M-D+1},\cdots,\bm{e}_{M}$ (or equivalently $D$ array steering vector $\bm{a}(\theta_{1}),\cdots,\bm{a}(\theta_{D})$).

To solve for the array steering vectors (thus AoA), MUSIC plots the reciprocal of squared distance $Q(\theta)$ for points along the $\theta$ continue to the noise subspace as a function of $\theta$:
\begin{equation}
Q(\theta)=\frac{1}{\bm{a}^{\mathrm{H}}(\theta)\bm{E}_{N}\bm{E}_{N}^{\mathrm{H}}\bm{a}(\theta)}
\end{equation}
This yields peaks in $Q(\theta)$ at the bearing of incident signals. It is similar to apply MUSIC algorithm for AoD spectrum estimation.

The following function \lstinline{naive_aoa} intends to estimate the 3D AoA based on the phase difference, which is similar to \eq{eqn:naive-aoa}. Note that the following algorithm only considers one path, and thus cannot be applied to mutlipath signals.

\begin{lstlisting}[language=MATLAB]
function [aoa_mat] = naive_aoa(csi_data, antenna_loc, est_rco)
    % naive_aoa
    % Input:
    %   - csi_data is the CSI used for angle estimation; [T S A E]
    %   - antenna_loc is the antenna location arrangement with the first antenna as a reference; [3 A]
    %   - est_rco is the estimated radio chain offset; [A 1]
    % Output:
    %   - aoa_mat is the angle estimation result; [3 T]

    global subcarrier_lambda;
    [packet_num, subcarrier_num, antenna_num, extra_num] = size(csi_data);
    csi_phase = unwrap(angle(csi_data), [], 2);    % [T S A E]
    % Get the antenna vector and its length.
    ant_diff = antenna_loc(:, 2:end) - antenna_loc(:, 1); % [3 A-1]
    ant_diff_length = vecnorm(ant_diff); % [1 A-1]
    ant_diff_normalize = ant_diff ./ ant_diff_length; % [3 A-1]
    % Calculate the phase difference.
    phase_diff = csi_phase(:, :, 2:end, :) - csi_phase(:, :, 1, :) - permute(est_rco(2:end, :), [4 3 1 2]); % [T S A-1 E]
    phase_diff = unwrap(phase_diff, [], 2);
    phase_diff = mod(phase_diff + pi, 2 * pi) - pi;
    % Broadcasting is performed, get the value of cos(theta) for each packet and each antenna pair.
    cos_mat = subcarrier_lambda .* phase_diff ./ (2 .* pi .* permute(ant_diff_length, [3 1 2])); % [T S A-1 E]
    cos_mat_mean = squeeze(mean(cos_mat, [2 4])); % [T A-1]
    % Symbolic nonlinear optimization are performed.
    syms x y
    % aoa_sol = [x;y;(1-sqrt(x^2 + y^2))];
    aoa_init = [sqrt(1/3);sqrt(1/3);sqrt(1/3)];
    aoa_mat_sol = zeros(3, packet_num);
    options = optimoptions('lsqnonlin', 'Algorithm', 'levenberg-marquardt', 'Display', 'none');
    parfor p = 1:packet_num
       cur_nonlinear_func = @(aoa_sol)ant_diff_normalize' * aoa_sol - cos_mat_mean(p, :)';
       cur_aoa_sol = lsqnonlin(cur_nonlinear_func, aoa_init, [], [], options);
       aoa_mat_sol(:, p) = cur_aoa_sol;
    end
    aoa_mat = aoa_mat_sol ./ vecnorm(aoa_mat_sol); % [3 T]
end
\end{lstlisting}

\subsection{Phase Shift Spectrum}
\label{sec:DFS}

\begin{figure}[h]
\centering
\includegraphics[width=.95\textwidth]{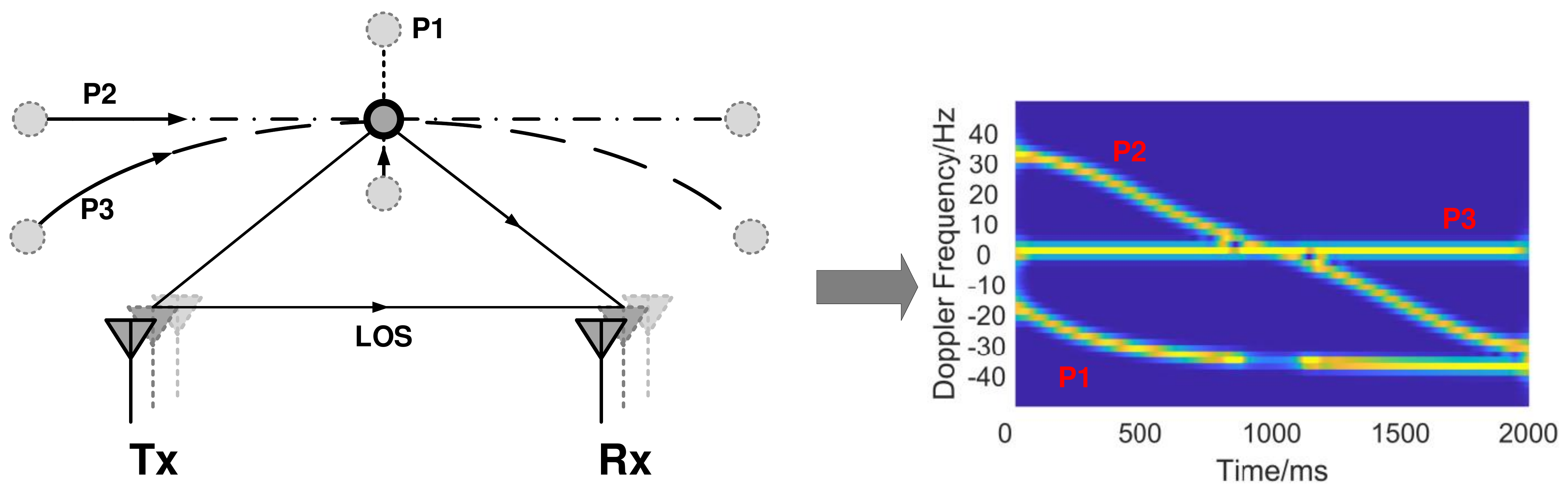}
\caption{Phase shift spectrum (or Doppler spectrum) of three different moving path.}
\label{fig:p2c2-dfs}
\end{figure}

Non-zero phase shift $\Delta \phi$ across different packets is caused by the relative movement of the transmitter, receiver, or objects in the propagation path of the signal. It equals the changing rate of the path length of the signal.  
When multiple packets are received in sequence, the CSI corresponding to the $i^{th}$ received packet is:
\begin{equation}
    H(i) = \sum_{n=1}^{N}\alpha_{n}(i)e^{j\phi_{n}(i)}, 
    \label{eq:p2c2_dfs}
\end{equation}
where $\phi_{n}$ is the phase of the $n^{th}$ path~\cite{chi2022widrone}.
Extract the phase of the the $n^{th}$ path in packet $i$ and $i+1$ respectively, and calculate the phase shift as:
\begin{equation}
    \Delta \phi_{n}(i) = \phi_{n}(i+1) - \phi_{n}(i).
    \label{eq:phase_diff}
\end{equation}

Intuitively, the phase difference indicates the distance change of the $n^{th}$ path between two consecutive packets: $\Delta d_{n}(i) = \frac{\Delta \phi_{n}(i)}{2 \pi} \lambda$.

Take a step further, and apply the short-time Fourier transform (STFT) within a sliding window, we can get the spectrum as shown in Figure~\ref{fig:p2c2-dfs}.
The frequency axis reveals the change rate of consecutive CSI data, and implicitly contains the path length change rate. 
Figure~\ref{fig:p2c2-dfs} demonstrates the phase shift spectrum (or the Doppler Spectrum) for three different moving paths.

The following function \lstinline{naive_stft} calculates the short-time Fourier transform of a series of CSI data. The generated spectrum can be used effectively for many wireless sensing tasks, like gesture recognition and fall detection. 

\begin{lstlisting}[language=MATLAB]
function stft_mat = naive_spectrum(csi_data, sample_rate, visable)
    % naive_spectrum
    % Input:
    %   - csi_data is the CSI used for STFT spectrum generation; [T S A L]
    %   - sample_rate determines the resolution of the time-domain and
    %   frequency-domain;
    % Output:
    %   - stft_mat is the generated STFT spectrum; [sample_rate/2 T]

    % Conjugate multiplication.
    csi_data = mean(csi_data .* conj(csi_data), [2 3 4]);
    % Calculate the STFT and visualization.
    stft_mat = stft(csi_data, sample_rate);
    % Visualization (optional).
    if visable
        stft(csi_data, sample_rate);
    end
end
\end{lstlisting}

\subsection{Body-coordinate Velocity Profile}

The limitation of the aforementioned spectrum is that, even the spectrum corresponding to the same activity will be different when the user moves at different locations or orientations relative to the Wi-Fi links.
To resolve this problem, Widar3.0~\cite{Widar3.0} proposes a domain-independent signal feature BVP (body-coordinate velocity profile) to characterize human activities.

\begin{figure}[tb]
\centering
\hspace{0.1in}\includegraphics[width=0.49\textwidth]{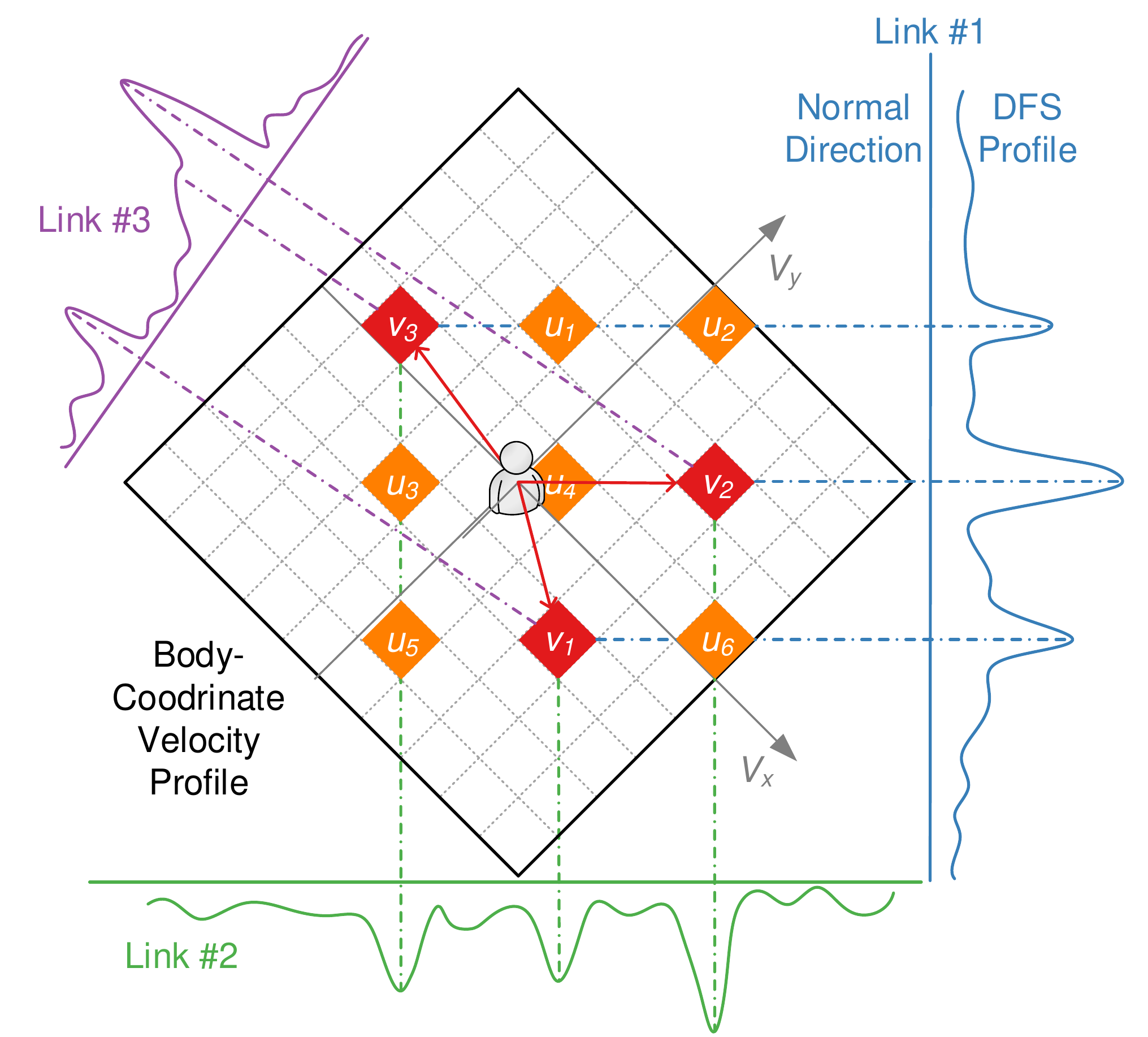}
\caption{Relationship between the BVP and Doppler spectrum. Each velocity component in BVP is projected onto the normal direction of a link, and contributes to the power of the corresponding radial velocity component in the Doppler spectrum.}
\label{fig:vp}
\end{figure}

The basic idea of BVP is shown in \fig{fig:vp}.
A BVP $\bm{V}$ is quantized as a discrete matrix with dimension as velocity components decomposed along each axis of the body coordinates.
For convenience, we establish the local body coordinates whose origin is the location of the person and positive x-axis aligns with the orientation of the person.
The person's location and orientation should be provided manually.
Currently, it is assumed that the global location and orientation of the person are available.
Then the known global locations of wireless transceivers can be transformed into the local body coordinates.
Thus, for better clarity, all locations and orientations used in the following derivation are in the local body coordinates.
Suppose the locations of the transmitter and the receiver of the $i^{th}$ link are $\vec{l}_{t}^{(i)}=(x_{t}^{(i)},y_{t}^{(i)})$, $\vec{l}_{r}^{(i)}=(x_{r}^{(i)},y_{r}^{(i)})$, respectively, then any velocity components $\vec{v}=(v_x, v_y)$ around the human body (\ie, the origin) will contribute its signal power to some frequency component, denoted as $f^{(i)}(\vec{v})$, in the Doppler spectrum of the $i^{th}$ link~\cite{qian2017widar}:
\begin{equation}
\label{eq:vp_relation}
f^{(i)}(\vec{v})=a_{x}^{(i)}v_x+a_{y}^{(i)}v_y.
\end{equation}
$a_{x}^{(i)}$ and $a_{y}^{(i)}$ are coefficients determined by locations of the transmitter and the receiver:
\begin{equation}
\label{eq:val_a}
\begin{aligned}
a_{x}^{(i)}=\frac{1}{\lambda}(\frac{x_{t}^{(i)}}{\|\vec{l}_{t}^{(i)}\|_2}+\frac{x_{r}^{(i)}}{\|\vec{l}_{r}^{(i)}\|_2}),\\
a_{y}^{(i)}=\frac{1}{\lambda}(\frac{y_{t}^{(i)}}{\|\vec{l}_{t}^{(i)}\|_2}+\frac{y_{r}^{(i)}}{\|\vec{l}_{r}^{(i)}\|_2}),
\end{aligned}
\end{equation}
where $\lambda$ is the wavelength of Wi-Fi signal.
As static components with zero Doppler spectrum (\eg, the line of sight signals and dominant reflections from static objects) are filtered out before the Doppler spectrum are calculated, only signals reflected by the person are retained.
Besides, when the person is close to the Wi-Fi link, only signals with one-time reflection have prominent magnitudes \cite{qian2018widar2}.
Thus, \eq{eq:vp_relation} holds valid for the gesture recognition scenario.
From the geometric view, \eq{eq:vp_relation} means that the 2-D velocity vector $\vec{v}$ is projected on a line whose direction vector is $\bm{d}^{(i)}=(-a_{y}^{(i)}, a_{x}^{(i)})$.
Suppose the person is on an ellipse curve whose foci are the transmitter and the receiver of the $i^{th}$ link, then $d^{(i)}$ is indeed the average direction of the ellipse at the person's location.
\fig{fig:vp} shows an example where the person generates three velocity components $\vec{v}_j, j=1,2,3$, and projection of the velocity components on the Doppler spectrum of three links. 

Since coefficients $a_{x}^{(i)}$ and $a_{y}^{(i)}$ only depend on the location of the $i^{th}$ link, the relation of projection of the BVP on the $i^{th}$ link is fixed.
Specifically, an assignment matrix $\bm{A}^{(i)}_{F\times N^2}$ can be defined:
\begin{equation}
\label{eq:ass}
A^{(i)}_{j,k}=\left\{\begin{array}{ll}
1 & f_j = f^{(i)}(\vec{v}_k) \\
0 & {\rm else}
\end{array}\right.,
\end{equation}
where $f_j$ is the $j^{th}$ frequency sampling point in the Doppler spectrum, and $\vec{v}_k$ is velocity component corresponding to the $k^{th}$ element of the vectorized BVP $\bm{V}$.
Thus, the relation between Doppler spectrum profile of the $i^{th}$ link and the BVP can be modeled as:
\begin{equation}
\label{eq:model}
\bm{D}^{(i)}=c^{(i)}\bm{A}^{(i)}\bm{V}
\end{equation}
where $c^{(i)}$ is the scaling factor due to propagation loss of the reflected signal.

Due to the sparsity of BVP, compressed sensing~\cite{donoho2006compressed} technique can be adopted to formulate the estimation of BVP as an $l_0$ optimization problem:
\begin{equation}
{\rm min}_{V}\sum_{i=1}^{M}|{\rm EMD}(\bm{A}^{(i)}\bm{V}, \bm{D}^{i})|+\eta\|V\|_0,
\end{equation}
where $M$ is the number of Wi-Fi links. 
The sparsity of the number of the velocity components is coerced by the term $\eta\|V\|_0$, where $\eta$ represents the sparsity coefficients and $\|\cdot\|_0$ is the number of non-zero velocity components.

${\rm EMD}(\cdot,\cdot)$ is the Earth Mover's Distance~\cite{rubner2001earth} between two distributions.
The selection of EMD rather than Euclidean distance is mainly due to two reasons.
First, the quantization of BVP introduces approximation error, \ie, projection of velocity components to the Doppler spectrum bin might be adjacent to the true one.
Such quantization error can be relieved by EMD, which takes the distance between bins into consideration.
Second, there are unknown scaling factors between the BVP and Doppler spectrum, making the Euclidean distance inapplicable.

\section{CSI Sanitization}

The wireless sensing models and features described in previous sections are consistent with the EM propagation theory and geometry. However, they don't consider the various types of noise caused by imperfect implementations of transceiver hardware~\cite{xie2018precise, xie2019md}.
This section focuses on various CSI error sources and the corresponding error cancellation algorithms.

For ease of illustration, code implementation for sanitization is provided, which is a \lstinline{main} function to call different error cancellation functions and test their performance.

\begin{lstlisting}[language=MATLAB]
%{
  CSI Sanitization Algorithm for Wi-Fi sensing.  
  - Input: csi data used for calibration, and csi data that need to be sanitized.
  - Output: sanitized csi data.

  To use this script, you need to:
  1. Make sure the csi data have been saved as .mat files.
  2. Check the .mat file to make sure they are in the correct form. 
  3. Set the parameters.
  
  Note that in this algorithm, the csi data should be a 4-D tensor with the size of [T S A L]:
  - T indicates the number of csi packets;
  - S indicates the number of subcarriers;
  - A indicates the number of antennas (i.e., the STS number in a MIMO system);
  - L indicates the number of HT-LTFs in a single PPDU;
  Say if we collect a 10 seconds csi data steam at 1 kHz sample rate (T = 10 * 1000), from a 3-antenna AP (A = 3),  with 802.11n standard (S = 57 subcarrier), without only one spatial stream (L = 1), the data size should be [10000 57 3 1].
%}

clear all;
addpath(genpath(pwd));

%% 0. Set parameters.
% Path of the calibration data;
calib_file_name = './data/csi_calib_test.mat';
% Path for storing the generated calibration templated.
calib_template_name = './data/calib_template_test.mat';
% Path of the raw CSI data.
src_file_name = './data/csi_src_test.mat';
% Path for storing the sanitized CSI data.
dst_file_name = './data/csi_dst_test.mat';

% Speed of light.
global c;
c = physconst('LightSpeed');
% Bandwidth.
global bw;
bw = 20e6;
% Subcarrier frequency.
global subcarrier_freq;
subcarrier_freq = linspace(5.8153e9, 5.8347e9, 57);
% Subcarrier wavelength.
global subcarrier_lambda;
subcarrier_lambda = c ./ subcarrier_freq;

% Antenna arrangement.
antenna_loc = [0, 0, 0; 0.0514665, 0, 0; 0, 0.0514665, 0]';
% Set the linear range of the CSI phase, which varies with NIC types.
linear_interval = (20:38)';

%% 1. Read the csi data for calibration and sanitization.
% Load the calibration data. 
csi_calib = load(calib_file_name).csi; % CSI for calibration.
% Load the raw CSI data.
csi_src = load(src_file_name).csi;      % Raw CSI.

%% 2. Choose different functions according to your task.
% Use cases:
% Make calibration template.
csi_calib_template = set_template(csi_calib, linear_interval, calib_template_name);
% Directly load the generated template.
csi_calib_template = load(calib_template_name).csi;
% Remove the nonlinear error.
csi_remove_nonlinear = nonlinear_calib(csi_src, csi_calib_template);
% Remove the STO (a.k.a SFO and PBD) by conjugate mulitplication.
csi_remove_sto = sto_calib_mul(csi_src);
% Remove the STO (a.k.a SFO and PBD) by conjugate division.
csi_remove_sto = sto_calib_div(csi_src);
% Estimate the CFO by frequency tracking.
est_cfo = cfo_calib(csi_src);
% Estimate the RCO.
est_rco = rco_calib(csi_calib);

%% 3. Save the sanitized data as needed.
csi = csi_remove_sto;
save(dst_file_name, 'csi');

%% 4. Perform various wireless sensing tasks.
% Test example 1: angle/direction estimation with imperfect CSI.
[packet_num, subcarrier_num, antenna_num, ~] = size(csi_src);
est_rco = rco_calib(csi_calib);
zero_rco = zeros(antenna_num, 1);
aoa_mat_error = naive_aoa(csi_src, antenna_loc, zero_rco);
aoa_mat = naive_aoa(csi_src, antenna_loc, est_rco);
aoa_gt = [0; 0; 1];
error_1 = mean(acos(aoa_gt' * aoa_mat_error));
error_2 = mean(acos(aoa_gt' * aoa_mat));
disp("Angle estimation error (in deg) without RCO removal: " + num2str(error_1));
disp("Angle estimation error (in deg) with RCO removal: " + num2str(error_2));

% Test example 2: intrusion detection with CSI.
csi_sto_calib = sto_calib_div(csi_src);
intrusion_flag_raw = naive_intrusion(csi_src, 3);
intrusion_flag_sanitized = naive_intrusion(csi_sto_calib, 3);
disp("Intrusion detection result without SFO/PDD removal: " + num2str(intrusion_flag_raw));
disp("Intrusion detection result with SFO/PDD removal: " + num2str(intrusion_flag_sanitized));
\end{lstlisting}

\subsection{Nonlinear Amplitude and Phase} 

The nonlinear amplitude and phase errors are caused by the imperfect analog domain filter implementation inside the hardware.
Specifically, it causes the extracted CSI amplitude and phase to be equivalently processed by a nonlinear function. Let $f(\cdot)$ and $g(\cdot)$ be the nonlinear modes of CSI amplitude and phase, respectively, the errorous CSI can be written as:
\begin{equation}
\tilde{H}(i,j,k) = \sum_{n = 1}^{N} {f(\alpha_{n}) e^{-jg(\phi_{n}(i,j,k))}} + N(i,j,k).
\end{equation}

Specifically, during the OFDM modulation process, each subcarrier of should have the same gain. In other words, the amplitude-frequency characteristic of CSI should be a horizontal straight line when a coaxial cable is used to connect the transceiver ports. However, actual measurements show that even without the multipath radio channel, there is still a similar "frequency selective fading" characteristic, \ie, the gain of each frequency band is different, showing an M-shaped amplitude-frequency characteristic curve. Similarly, when using a coaxial cable to connect the transceiver port, the CSI phase-frequency characteristics obtained by the NIC are not an ideal straight line with slope, but an S-shaped curve with certain nonlinearity.

After extensive research and experiments, we have observed two facts.
\begin{itemize}
\item First, for a specific type of NIC, the nonlinear amplitude/phase error of CSI is fixed. This means that the correction task can be accomplished if we use a known length coaxial cable connected to the transceiver port. Before performing the sensing task, measure a representative set of CSI amplitude and phase, record the nonlinear characteristics, and eliminate the nonlinearity in the subsequent steps.
\item Second, we observe that the middle part of the subcarrier is free of nonlinear errors, and the nonlinear characteristics of both sides are also fixed.
\end{itemize}

Therefore, the function below the code performs the following steps to tackle the CSI nonlinearity:
\begin{enumerate}
\item Read in a set of CSI data measured using a coaxial cable.
\item Get its amplitude and phase.
\item Normalize its amplitude and record it as the amplitude template.
\item Unwrap the phases, then perform a linear fit to the middle part of its subcarriers. Subtract the linear fit result to obtain the nonlinear phase error template.
\end{enumerate}

Finally, the nonlinear amplitude and nonlinear phase components are saved in the form of $\alpha e^{j\phi}$, which indicates the nonlinear error template of a specific type of NIC.

\begin{lstlisting}[language=MATLAB]
function [csi_calib_template] = set_template(csi_calib, linear_interval, calib_template_name)
    % set_template
    % Input:
    %   - csi_calib is the reference csi at given distance and angle; [T S A L]
    %   - linear_interval is the linear range of the csi phase, which varies across different types of NICs;
    %   - calib_template_name is the saved path of the generated template;
    % Output:
    %   - csi_calib_template is the generated template for csi calibration; [1 S A L]

    [packet_num, subcarrier_num, antenna_num, extra_num] = size(csi_calib);
    csi_amp = abs(csi_calib);                       % [T S A L]
    csi_phase = unwrap(angle(csi_calib), [], 2);    % [T S A L]
    csi_amp_template = mean(csi_amp ./ mean(csi_amp, 2), 1); % [1 S A L]
    nonlinear_phase_error = zeros(size(csi_calib));          % [T S A L]
    for p = 1:packet_num
        for a = 1:antenna_num
            for e = 1:extra_num
                linear_model = fit(linear_interval, squeeze(csi_phase(p, linear_interval, a, e))', 'poly1');
                nonlinear_phase_error(p, :, a, e) = csi_phase(p, :, a, e) - linear_model(1:subcarrier_num)';
            end
        end
    end
    csi_phase_template = mean(nonlinear_phase_error, 1); % [1 S A L]
    csi_phase_template(1, linear_interval, :, :) = 0;
    csi_calib_template = csi_amp_template .* exp(1i * csi_phase_template); % [1 S A L]
    csi = csi_calib_template;
    save(calib_template_name, 'csi'); % [1 S A L]
end
\end{lstlisting}

After getting the nonlinear error template, the sanitization process begins.
For the raw CSI data \lstinline{csi_data} collected in real time, we divide it by the error template \lstinline{csi_calib}.
This operation is equivalent to ``dividing the original amplitude by the normalized nonlinear amplitude'' and ``subtracting the original phase from the nonlinear phase'', so that the returned CSI data \lstinline{csi_proc} has sanitized amplitude and phase.

\begin{lstlisting}[language=MATLAB]
function [csi_remove_nonlinear] = nonlinear_calib(csi_calib, csi_calib_template)
    % nonlinear_calib
    % Input:
    %   - csi_src is the raw csi which needs to be calibrated; [T S A L]
    %   - csi_calib_template is the reference csi for calibration; [1 S A L]
    % Output:
    %   - csi_remove_nonlinear is the csi data in which the nonlinear error has been eliminated; [T S A L]

    csi_amp = abs(csi_calib);                       % [T S A L]
    csi_phase = unwrap(angle(csi_calib), [], 2);    % [T S A L]
    csi_unwrap = csi_amp .* exp(1i * csi_phase);     % [T S A L]
    % Broadcasting is performed.
    csi_remove_nonlinear = csi_unwrap ./ csi_calib_template;
end
\end{lstlisting}

\subsection{Automatic Gain Control Uncertainty} 
Automatic gain control (AGC) induces a random gain $\beta$ in each received CSI packet.
\begin{equation}
\tilde{H}(i,j,k) = \sum_{n = 1}^{N} {\beta_{i} \alpha_{n} e^{-j\phi_{n}(i,j,k)}} + N(i,j,k). 
\end{equation}

There are two ways to eliminate the AGC error: 1) Disable the AGC function of the wireless driver; 2) Compensate the amplitude of the measured CSI based on the reported AGC.

\begin{lstlisting}[language=MATLAB]
function [csi_remove_agc] = agc_calib(csi_src, csi_agc)
    % rco_calib
    % Input:
    %   - csi_src is the raw csi which needs to be calibrated; [T S A L]
    %   - csi_agc is the AGC amplitude reported by the NIC; [T, 1]
    % Output:
    %   - csi_remove_agc is the csi data in which the AGC uncertainty has been eliminated; [T S A L]

    % Broadcasting is performed.
    csi_remove_agc = csi_src ./ csi_agc;
end
\end{lstlisting}

\subsection{Radio Chain Offset} 
Radio chain offset (RCO) is the random phase variation $\epsilon_{\phi}$ introduced between different Tx/Rx chains (transceiver antenna pairs). The RCO is reset each time the NIC is powered up.
\begin{equation}
    \tilde{\phi}_{n}(i,j,k) = 2\pi(f_c + \Delta f_j + f_{D}) \tau_n(i, j, k) + \epsilon_{\phi}, 
\end{equation}
RCO induces a biast of the phase-frequency characteristic curve.
It undermines the accuracy of features such as AoA or ToF. Fortunately, we found that this type of phase deviation is consistent between each successive packet sent and therefore doesn't affect the performance of temporal tracking or sensing, and that this type of error can be eliminated by the following steps:
\begin{enumerate}
    \item Once the NIC power-up, connect the transceiver port using a coaxial cable of known length and record the phase information \lstinline{calib_phase}.
    \item during subsequent measurements, subtracting this phase information from the measured phase \lstinline{csi_phase}.
\end{enumerate}

\begin{lstlisting}[language=MATLAB]
function [est_rco] = rco_calib(csi_calib)
    % rco_calib
    % Input:
    %   - csi_calib is the reference csi at given distance and angle; [T S A L]
    % Output:
    %   - est_rco is the estimated RCO; [A 1]

    antenna_num = size(csi_calib, 3);
    csi_phase = unwrap(angle(csi_calib), [], 1);    % [T S A L]
    avg_phase = zeros(antenna_num, 1);
    for a = 1:antenna_num
        avg_phase(a, 1) = mean(csi_phase(:, :, a, 1), 'all');
    end
    est_rco = avg_phase - avg_phase(1, 1);
end
\end{lstlisting}

\subsection{Central Frequency Offset} 
Central frequency offset (CFO), which is caused by the frequency desynchronization on both sides of the transceiver, leads to random frequency shift $ \epsilon_{f}$ of each received CSI.
\begin{equation}
    \tilde{\phi}_{n}(i,j,k) = 2\pi(f_c + \Delta f_j + f_{D} + \epsilon_{f}) \tau_n(i, j, k) = \phi_{n}(i,j,k) + \epsilon_{f}\tau_n(i, j, k).
\end{equation}

The CFO induces an extra bias (\ie, an overall up and down shift) of the phase-frequency characteristic curve.

To eliminate CFO, we need to insert multiple HT-LTFs in the same PPDU (Wi-Fi data frame), and therefore obtaining multiple CSI measurements. Since the time interval between multiple HT-LTFs is strictly controlled to $4 \mu s$ according to the 802.11 protocol, the phase difference between two HT-LTFs is induced by the CFO within $\Delta t = 4\mu s$. Thus, the approximate value of the CFO can be recovered.

\begin{lstlisting}[language=MATLAB]
function [est_cfo] = cfo_calib(csi_src)
    % cfo_calib
    % Input:
    %   - csi_src is the csi data with two HT-LTFs; [T S A L]
    % Output:
    %   - est_cfo is the estimated frequency offset; 

    delta_time = 4e-6;
    phase_1 = angle(csi_src(:, :, :, 1));
    phase_2 = angle(csi_src(:, :, :, 2));
    phase_diff = mean(phase_2 - phase_1, 3); % [T S A 1]
    est_cfo = mean(phase_diff ./ delta_time, 2);
end
\end{lstlisting}

\subsection{Sampling Frequency Offset and Packet Detection Delay} 

The sampling frequency offset (SFO), which appears to be an error in frequency domain, are generally considered as an equivalent time shift due to frequency asynchrony. The packet detection delay (PDD) is a time delay. Therefore, despite their distinct causes, they are often discussed together as a ``time offset'' together.
\begin{equation}
    \tilde{\phi}_{n}(i,j,k) = 2\pi(f_c + \Delta f_j + f_{D}) (\tau_n(i, j, k) + \epsilon_{t}) = \phi_{n}(i,j,k) + 2\pi(f_c + \Delta f_j + f_{D}) \epsilon_{t}.
\end{equation}

This type of error is critical because the time delay can be confused with real ToF and thus affect the accuracy of the ranging accuarcy. Specifically, this deviation will be characterized in the phase-frequency characteristic curve as a change in slope, since this time deviation causes different phase changes with different sub-bands $\Delta f_j$.

Currently, there is no ``perfect algorithm'' to solve this type of error. Conjugate multiplication and division are the only two methods to eliminate the SFO and PDD. 
The code of both of them are listed below. By appling the conjugate multiplication or division, the $\epsilon_{t}$ is eliminated, at the cost of losing absolute ToF measurement.

\begin{lstlisting}[language=MATLAB]
function [csi_remove_sto] = sto_calib_mul(csi_src)
    % sto_calib_mul
    % Input:
    %   - csi_src is the csi data with sto; [T S A L]
    % Output:
    %   - csi_remove_sto is the csi data without sto; [T S A L]

    antenna_num = size(csi_src, 3);
    csi_remove_sto = zeros(size(csi_src));
    for a = 1:antenna_num
        a_nxt = mod(a, antenna_num) + 1;
        csi_remove_sto(:, :, a, :) = csi_src(:, :, a, :) .* conj(csi_src(:, :, a_nxt, :));
    end
end
\end{lstlisting}

\begin{lstlisting}[language=MATLAB]
function [csi_remove_sto] = sto_calib_div(csi_src)
    % sto_calib_div
    % Input:
    %   - csi_src is the csi data with sto; [T S A L]
    % Output:
    %   - csi_remove_sto is the csi data without sto; [T S A L]

    antenna_num = size(csi_src, 3);
    csi_remove_sto = zeros(size(csi_src));
    for a = 1:antenna_num
        a_nxt = mod(a, antenna_num) + 1;
        csi_remove_sto(:, :, a, :) = csi_src(:, :, a, :) ./ csi_src(:, :, a_nxt, :);
    end
end
\end{lstlisting}

To sum up, there are various forms of errors in Wi-Fi CSI measurements, including fixed bias and random errors. Each of them have different impacts on the localization, tracking, and sensing tasks. 
The erroneous CSI form can be finally written as:
\begin{equation}
\tilde{H}(i,j,k) = \sum_{n = 1}^{N} {\beta_{i}f(\alpha_{n}) e^{-jg(\phi_{n}(i,j,k))}} + N(i,j,k), \\
\tilde{\phi}_{n}(i,j,k) = 2\pi(f_c + \Delta f_j + f_{D} + \epsilon_{f}) (\tau_n(i, j, k) + \epsilon_{t}) + \epsilon_{\phi}.
\end{equation}
\section{Wireless Sensing with Deep Learning}

This section introduces a series of learning algorithms, especially the prevalent deep neural network models such as CNN and RNN, and their applications in wireless sensing.
This section also proposes a complex-valued neural network to accomplish learning and inference based on wireless features efficiently.

\subsection{Convolutional Neural Network}
\label{sec:cnn}

\begin{figure}[tb]
\centering
\hspace{0.1in}\includegraphics[width=0.49\textwidth]{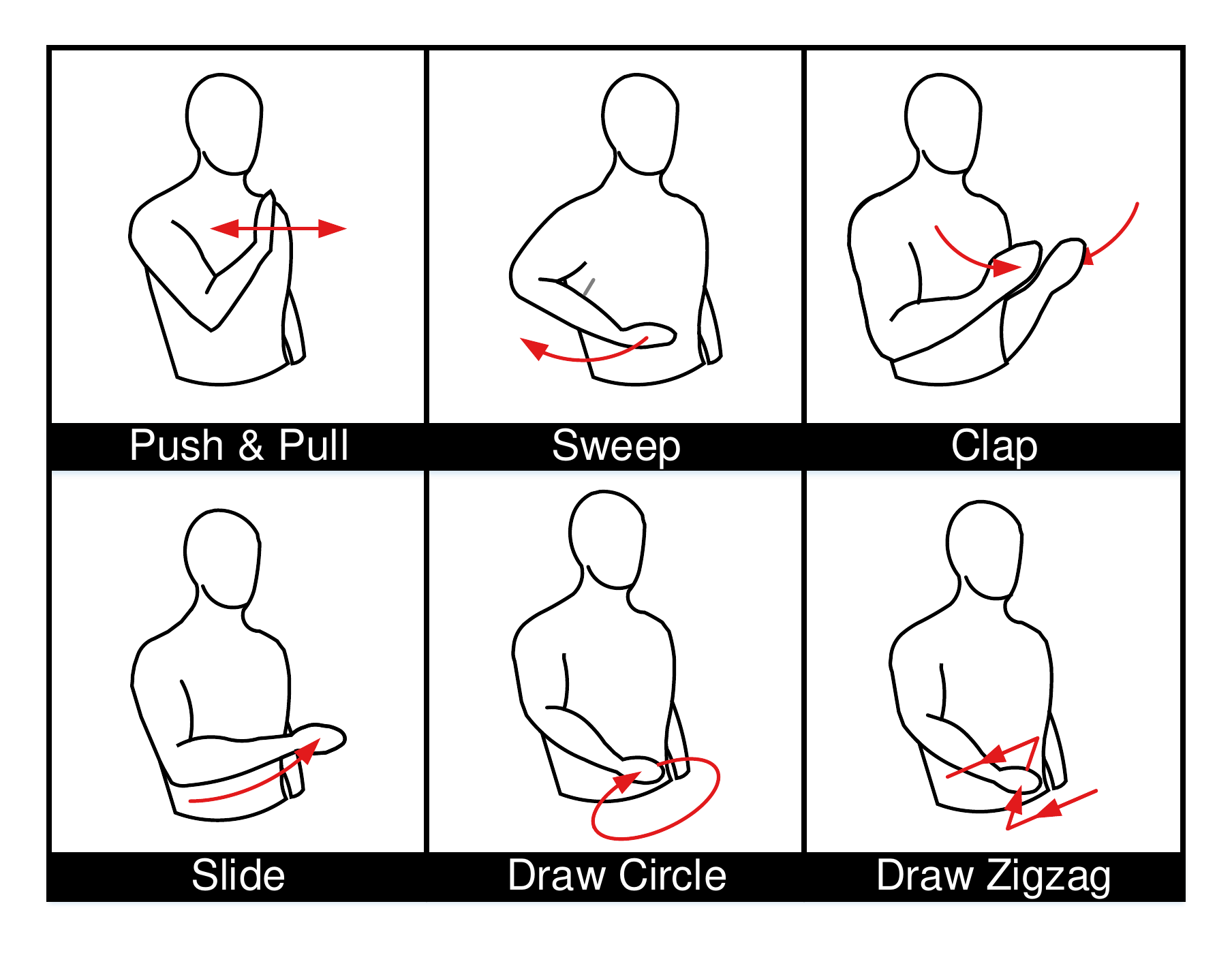}
\caption{Sketches of gestures evaluated in the experiment.}
\label{fig:gestures}
\end{figure}

\begin{figure}[tb]
\centering
\hspace{0.1in}\includegraphics[width=0.49\textwidth]{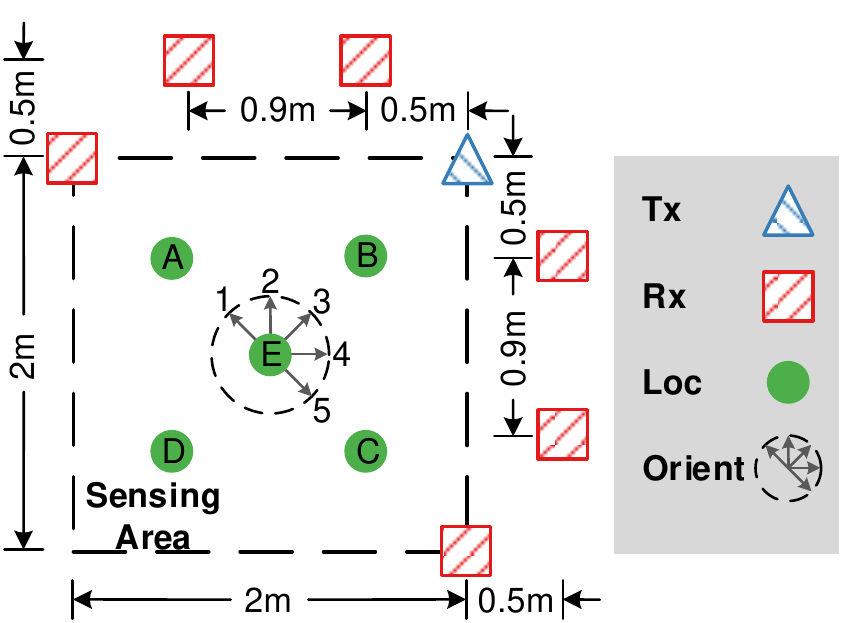}
\caption{The setup of WiFi devices for gesture recognition task.}
\label{fig:device_layout}
\end{figure}

\begin{figure}[tb]
\centering
\hspace{0.1in}\includegraphics[width=0.8\textwidth]{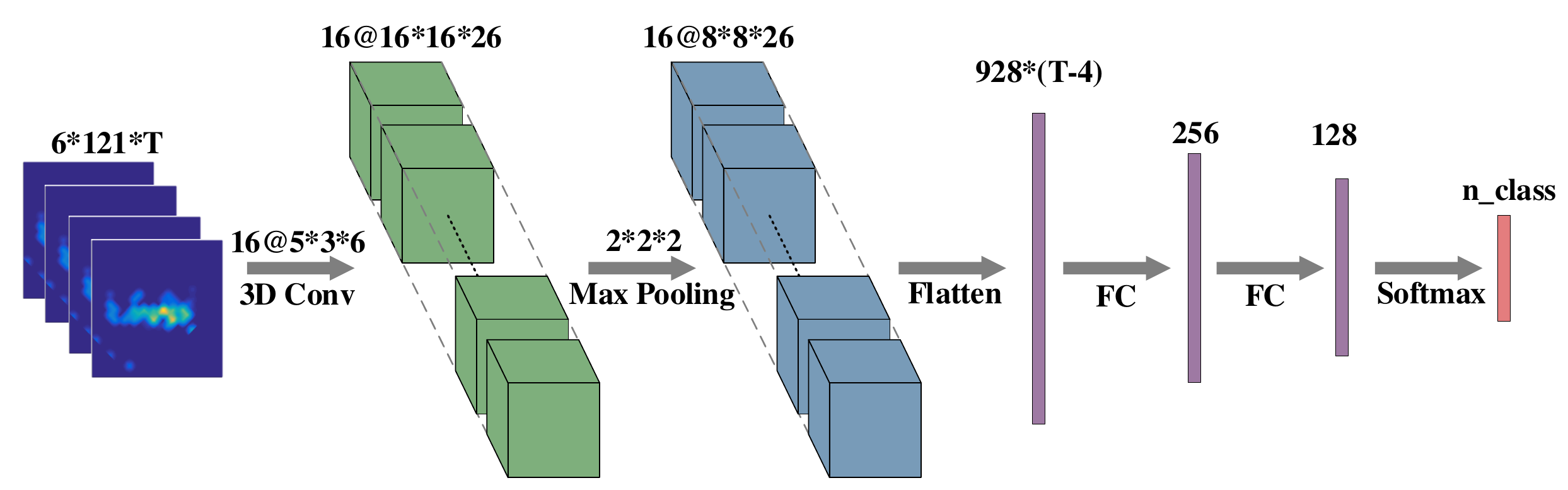}
\caption{Convolutional Neural Network architecture.}
\label{fig:cnn_arch}
\end{figure}

Convolutional Neural Network (CNN) contributes to the recent advances in understanding images, videos, and audios.
Some works~\cite{RFPose3D,Widar3.0,WiPose} have exploited CNN for wireless signal understanding in wireless sensing tasks and achieved promising performance.
This section will present a working example to demonstrate how to apply CNN for wireless sensing.
Specifically, we use commodity Wi-Fi to recognize six human gestures.
The gestures are illustrated in \fig{fig:gestures}.
We deploy a Wi-Fi transmitter and six receivers in a typical classroom, and the device setup is sketched in \fig{fig:device_layout}.
The users are asked to perform gestures at the five marked locations and to five orientations.
The data samples can be found in our released dataset~\cite{widar3_dataset}.
We extract DFS from raw CSI signals and feed them into a CNN network.
The network architecture is shown in \fig{fig:cnn_arch}.

We now introduce the implementation code in detail.

First, some necessary packages are imported.
We use Keras~\cite{chollet2015keras} API with TensorFlow as the backend to demonstrate how to implement the neural network.

\begin{lstlisting}[language=Python]
import os,sys
import numpy as np
import scipy.io as scio
import tensorflow as tf
import keras
from keras.layers import Input, GRU, Dense, Flatten, Dropout, Conv2D, Conv3D, MaxPooling2D, MaxPooling3D, TimeDistributed, Bidirectional, Multiply, Permute, RepeatVector, Concatenate, Dot, Lambda
from keras.models import Model, load_model
import keras.backend as K
from sklearn.metrics import confusion_matrix
from keras.backend.tensorflow_backend import set_session
from sklearn.model_selection import train_test_split
\end{lstlisting}

Then we define some parameters, including the hyperparameters and the data path.
The fraction of testing data is defined as 0.1.
To simplify the problem, we only use six gesture types in the widar3.0 dataset.
\begin{lstlisting}[language=Python]
# Parameters
fraction_for_test = 0.1
data_dir = 'widar30dataset/DFS/20181130/'
ALL_MOTION = [1,2,3,4,5,6]
N_MOTION = len(ALL_MOTION)
T_MAX = 0
n_epochs = 200
f_dropout_ratio = 0.5
n_gru_hidden_units = 64
n_batch_size = 32
f_learning_rate = 0.001
\end{lstlisting}

The program begins with loading data with the predefined function \lstinline{load_data}.
The loaded data are split into train and test by calling the API function \lstinline{train_test_split}.
The labels of the training data are encoded into the one-hot format with the predefined function \lstinline{onehot_encoding}.

\begin{lstlisting}[language=Python]
# Load data
data, label = load_data(data_dir)
print('\nLoaded dataset of ' + str(label.shape[0]) + ' samples, each sized ' + str(data[0,:,:,:,:].shape) + '\n')

# Split train and test
[data_train, data_test, label_train, label_test] = train_test_split(data, label, test_size=fraction_for_test)
print('\nTrain on ' + str(label_train.shape[0]) + ' samples\n' +\
    'Test on ' + str(label_test.shape[0]) + ' samples\n')

# One-hot encoding for train data
label_train = onehot_encoding(label_train, N_MOTION)
\end{lstlisting}

After loading and formatting the training and testing data, we defined the model with the predefined function \lstinline{build_model}.
After that, we train the model by calling the API function \lstinline{fit}.
The input data and label are specified in the parameters.
The fraction of validation data is specified as 0.1.
\begin{lstlisting}[language=Python]
# Train Model
model = build_model(input_shape=(T_MAX, 6, 121, 1), n_class=N_MOTION)
model.summary()
model.fit({'name_model_input': data_train},{'name_model_output': label_train},
        batch_size=n_batch_size,
        epochs=n_epochs,
        verbose=1,
        validation_split=0.1, shuffle=True)
\end{lstlisting}

After the training process, we evaluate the model with the test dataset.
The predictions are converted from one-hot format to integers and are used to calculate the confusion matrix and accuracy.

\begin{lstlisting}[language=Python]
# Testing...
print('Testing...')
label_test_pred = model.predict(data_test)
label_test_pred = np.argmax(label_test_pred, axis = -1) + 1

# Confusion Matrix
cm = confusion_matrix(label_test, label_test_pred)
print(cm)
cm = cm.astype('float')/cm.sum(axis=1)[:, np.newaxis]
cm = np.around(cm, decimals=2)
print(cm)

# Accuracy
test_accuracy = np.sum(label_test == label_test_pred) / (label_test.shape[0])
print(test_accuracy)
\end{lstlisting}

The predefined \lstinline{onehot_encoding} function convert the label to one-hot format.
\begin{lstlisting}[language=Python]
def onehot_encoding(label, num_class):
    # label(ndarray)=>_label(ndarray): [N,]=>[N,num_class]
    label = np.array(label).astype('int32')
    label = np.squeeze(label)
    _label = np.eye(num_class)[label-1]
    return _label
\end{lstlisting}

The predefined \lstinline{load_data} function is used to load all data samples and labels from a directory.
Each file in the directory corresponds to a single data sample.
Each data sample is normalized with the predefined \lstinline{normalize_data} function.
It is worth noting that the data samples have different time durations.
We use a predefined \lstinline{zero_padding} function to make their durations the same as the longest one.

\begin{lstlisting}[language=Python]
def load_data(path_to_data):
    global T_MAX
    data = []
    label = []
    for data_root, data_dirs, data_files in os.walk(path_to_data):
        for data_file_name in data_files:

            file_path = os.path.join(data_root,data_file_name)
            try:
                data_1 = scio.loadmat(file_path)['doppler_spectrum']    # [6,121,T]
                label_1 = int(data_file_name.split('-')[1])
                location = int(data_file_name.split('-')[2])
                orientation = int(data_file_name.split('-')[3])
                repetition = int(data_file_name.split('-')[4])
                
                # Downsample
                data_1 = data_1[:,:,0::10]

                # Select Motion
                if (label_1 not in ALL_MOTION):
                    continue

                # Normalization
                data_normed_1 = normalize_data(data_1)
                
                # Update T_MAX
                if T_MAX < np.array(data_1).shape[2]:
                    T_MAX = np.array(data_1).shape[2]                
            except Exception:
                continue

            # Save List
            data.append(data_normed_1.tolist())
            label.append(label_1)
            
    # Zero-padding
    data = zero_padding(data, T_MAX)

    # Swap axes
    data = np.swapaxes(np.swapaxes(data, 1, 3), 2, 3)   # [N,6,121,T_MAX]=>[N,T_MAX,6,121]
    data = np.expand_dims(data, axis=-1)                # [N,T_MAX,6,121]=>[N,T_MAX,6,121,1]

    # Convert label to ndarray
    label = np.array(label)

    # data(ndarray): [N,T_MAX,6,121,1], label(ndarray): [N,]
    return data, label
\end{lstlisting}

The \lstinline{normalize_data} function is used to normalize the loaded data samples.
Each data sample has a dimension of $[6, 121, T]$, in which the number "6" represents the number of Wi-Fi receivers, the number "121" represents the frequency bins, and the "T" represents the time durations.
To normalize a sample, we scale the data to be in the range of $[0, 1]$ for each time snapshot.
\begin{lstlisting}[language=Python]
def normalize_data(data_1):
    # data(ndarray)=>data_norm(ndarray): [6,121,T]=>[6,121,T]
    data_1_max = np.amax(data_1,(0,1),keepdims=True)    # [6,121,T]=>[1,1,T]
    data_1_min = np.amin(data_1,(0,1),keepdims=True)    # [6,121,T]=>[1,1,T]
    data_1_max_rep = np.tile(data_1_max,(data_1.shape[0],data_1.shape[1],1))    # [1,1,T]=>[6,121,T]
    data_1_min_rep = np.tile(data_1_min,(data_1.shape[0],data_1.shape[1],1))    # [1,1,T]=>[6,121,T]
    data_1_norm = (data_1 - data_1_min_rep) / (data_1_max_rep - data_1_min_rep + sys.float_info.min)
    return  data_1_norm
\end{lstlisting}

The \lstinline{zero_padding} function is used to align all the data samples to have the same duration.
The padded length is specified by the parameter \lstinline{T_MAX}.
\begin{lstlisting}[language=Python]
def zero_padding(data, T_MAX):
    # data(list)=>data_pad(ndarray): [6,121,T1/T2/...]=>[6,121,T_MAX]
    data_pad = []
    for i in range(len(data)):
        t = np.array(data[i]).shape[2]
        data_pad.append(np.pad(data[i], ((0,0),(0,0),(T_MAX - t,0)), 'constant', constant_values = 0).tolist())
    return np.array(data_pad)
\end{lstlisting}

In this function, we define the network structure.
The input layer is specified with the API function \lstinline{Input}, which has the parameters to define the input shape, data type, and the layer name.
Following the input layer, we use a three-dimensional convolutional layer, a max-pooling layer, and two fully connected layers.
The output layer is specified with the API function \lstinline{Output}, which has the parameters to define the activation function, the dimension, and the name.
At last, we finalize the model with the API function \lstinline{Model} and \lstinline{compile}.
The optimizer is specified to \lstinline{RMSprop}, and the loss is specified to \lstinline{categorical_crossentropy}.

\begin{lstlisting}[language=Python]
def build_model(input_shape, n_class):
    model_input = Input(shape=input_shape, dtype='float32', name='name_model_input')    # (@,T_MAX,6,121,1)

    # CNN
    x = Conv3D(16,kernel_size=(5,3,6),activation='relu',data_format='channels_last',\
        input_shape=input_shape)(model_input)   # (@,T_MAX-4,6,121,1)=>(@,T_MAX-4,4,116,16)
    x = MaxPooling3D(pool_size=(2,2,2))(x)      # (@,T_MAX-4,4,116,16)=>(@,T_MAX-4,2,58,16)

    x = Flatten()(x)                            # (@,T_MAX-4,2,58,16)=>(@,(T_MAX-4)*2*58*16)
    x = Dense(256,activation='relu')(x)         # (@,(T_MAX-4)*2*58*16)=>(@,256)
    x = Dropout(f_dropout_ratio)(x)
    x = Dense(128,activation='relu')(x)         # (@,256)=>(@,128)
    x = Dropout(f_dropout_ratio)(x)
    model_output = Dense(n_class, activation='softmax', name='name_model_output')(x)  # (@,128)=>(@,n_class) 

    # Model compiling
    model = Model(inputs=model_input, outputs=model_output)
    model.compile(optimizer=keras.optimizers.RMSprop(lr=f_learning_rate),
                    loss='categorical_crossentropy',
                    metrics=['accuracy']
                )
    return model
\end{lstlisting}

\subsection{Recurrent Neural Network}
\label{sec:rnn}

\begin{figure}[tb]
\centering
\hspace{0.1in}\includegraphics[width=0.9\textwidth]{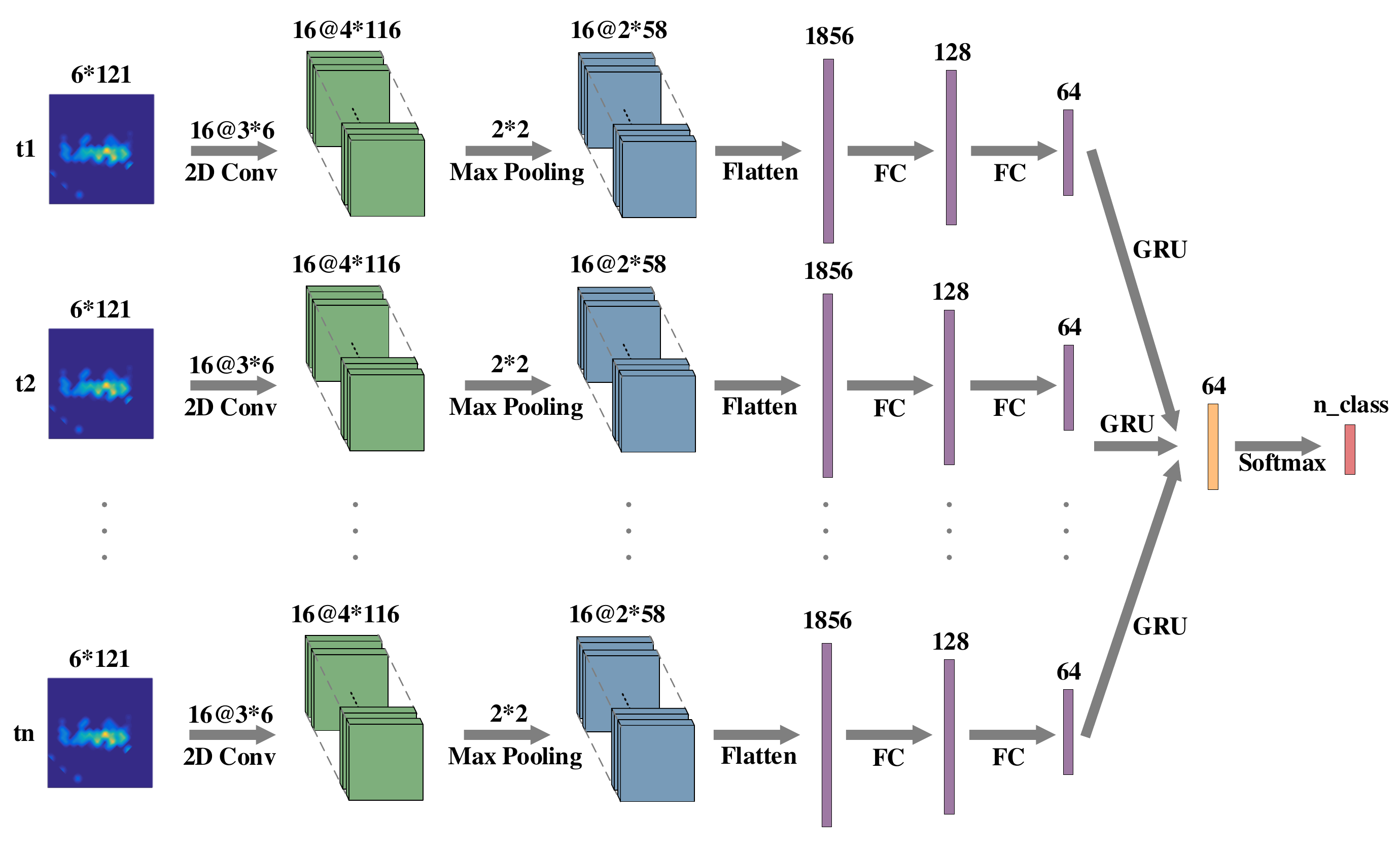}
\caption{Recurrent Neural Network architecture.}
\label{fig:rnn_arch}
\end{figure}

Recurrent Neural Network (RNN) is designed for modeling temporal dynamics of sequences and is commonly used for time series data analysis like speech recognition and natural language processing.
Wireless signals are highly correlated over time and can be processed with RNN.
Some works~\cite{Widar3.0, WiPose} have demonstrated the potential of RNN for wireless sensing tasks.
In this section, we will present a working example of combining CNN and RNN to perform gesture recognition with Wi-Fi.
The experimental settings are the same as in \sect{sec:cnn}.
We also extract DFS from the raw CSI as the input feature of the network.
The network architecture is shown in \fig{fig:rnn_arch}.

We now introduce the implementation code in detail.

Most of the code is the same as in \sect{sec:cnn} except for the model definition.
To define the model, we use two-dimensional convolutional layer and max-pooling layer on the dimensional except for the time dimension of the data.
We adopt the GRU layer as the recurrent layer.

\begin{lstlisting}[language=Python]
def build_model(input_shape, n_class):
    model_input = Input(shape=input_shape, dtype='float32', name='name_model_input')    # (@,T_MAX,6,121,1)

    # CNN+RNN
    x = TimeDistributed(Conv2D(16,kernel_size=(3,6),activation='relu',data_format='channels_last',\
        input_shape=input_shape))(model_input)              # (@,T_MAX,6,121,1)=>(@,T_MAX,4,116,16)
    x = TimeDistributed(MaxPooling2D(pool_size=(2,2)))(x)   # (@,T_MAX,4,116,16)=>(@,T_MAX,2,58,16)

    x = TimeDistributed(Flatten())(x)                       # (@,T_MAX,2,58,16)=>(@,T_MAX,2*58*16)
    x = TimeDistributed(Dense(128,activation='relu'))(x)    # (@,T_MAX,2*58*16)=>(@,T_MAX,128)
    x = TimeDistributed(Dropout(f_dropout_ratio))(x)
    x = TimeDistributed(Dense(64,activation='relu'))(x)     # (@,T_MAX,128)=>(@,T_MAX,64)

    x = GRU(n_gru_hidden_units,return_sequences=False)(x)   # (@,T_MAX,64)=>(@,64)
    x = Dropout(f_dropout_ratio)(x)
    model_output = Dense(n_class, activation='softmax', name='name_model_output')(x)  # (@,64)=>(@,n_class) 

    # Model compiling
    model = Model(inputs=model_input, outputs=model_output)
    model.compile(optimizer=keras.optimizers.RMSprop(lr=f_learning_rate),
                    loss='categorical_crossentropy',
                    metrics=['accuracy']
                )
    return model
\end{lstlisting}

\subsection{Adversarial Learning}

\begin{figure}[tb]
\centering
\hspace{0.1in}\includegraphics[width=0.8\textwidth]{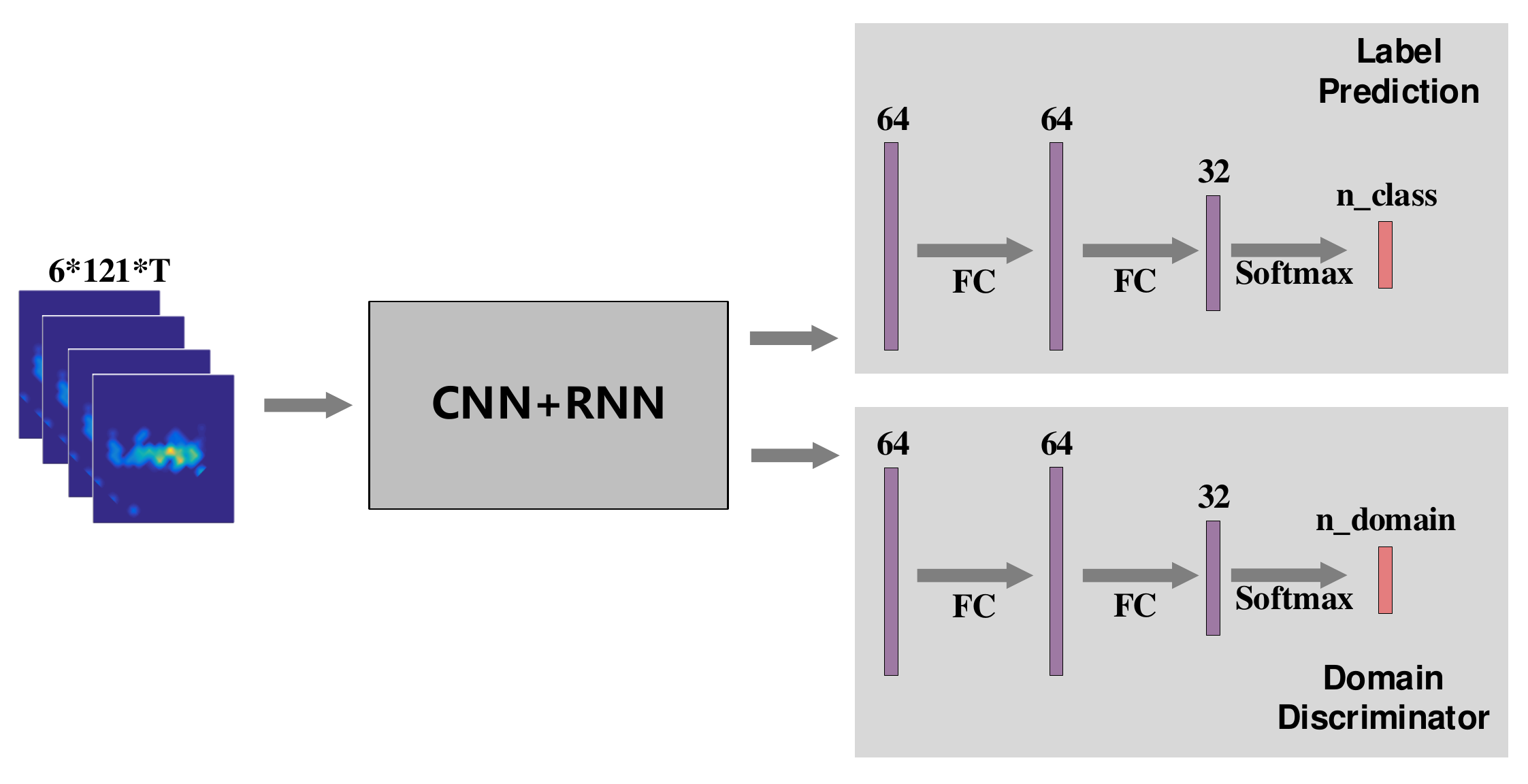}
\caption{Adversarial learning network architecture.}
\label{fig:adv_arch}
\end{figure}

Except for the basic neural network components, some high-level network architectures also play essential roles in wireless sensing.
Similar to computer vision tasks, wireless sensing also suffer from domain misalignment problem.
Wireless signals can be reflected by the surrounding objects during propagation and will be flooded with target-irrelevant signal components.
The sensing system trained in one deployment environment can hardly be applied directly in other settings without adaptation.
Some works~\cite{jiang2018towards} try to adopt adversarial learning techniques to tackle this problem and achieve promising performance.
This section will give an example of how to apply this technique in wireless sensing tasks.
Specifically, we build a gesture recognition system with Wi-Fi, similar to that in \sect{sec:cnn}.
We try to achieve consistent performance across different human locations and orientations.
The network architecture is shown in \fig{fig:adv_arch}.

We now introduce the implementation code in detail.

In the Widar3.0 dataset~\cite{widar3_dataset}, we collect gesture data when the users stand at different locations.
As discussed in \sect{sec:DFS}, human locations have significant impact on the DFS measurements.
To mitigate this impact, we treat human locations as different domains and build an adversarial learning network to recognize gestures irrespective of domains.
In the program, we first load data, labels, and domains from the dataset and split them into train and test.
Both label and domain are encoded into the one-hot format.

\begin{lstlisting}[language=Python]
# Load data
data, label, domain = load_data(data_dir)
print('\nLoaded dataset of ' + str(label.shape[0]) + ' samples, each sized ' + str(data[0,:,:,:,:].shape) + '\n')

# Split train and test
[data_train, data_test, label_train, label_test, domain_train, domain_test] = train_test_split(data, label, domain, test_size=fraction_for_test)
print('\nTrain on ' + str(label_train.shape[0]) + ' samples\n' +\
    'Test on ' + str(label_test.shape[0]) + ' samples\n')

# One-hot encoding for train data
label_train = onehot_encoding(label_train, N_MOTION)
domain_train = onehot_encoding(domain_train, N_LOCATION)
\end{lstlisting}

After loading and formating data, we built the network and trained it from scratch.
The training data, label, and domain are passed to the API function \lstinline{fit} for training.

\begin{lstlisting}[language=Python]
# Train Model
model = build_model(input_shape=(T_MAX, 6, 121, 1), n_class=N_MOTION, n_domain=N_LOCATION)
model.summary()
model.fit({'name_model_input': data_train},{'name_model_output_label': label_train, 'name_model_output_domain': domain_train},
        batch_size=n_batch_size,
        epochs=n_epochs,
        verbose=1,
        validation_split=0.1, shuffle=True)
\end{lstlisting}

After the training process finishes, we evaluate the network with the test samples.
Note that the adversarial network has both label and domain prediction outputs. We only use the label output for accuracy evaluation.

\begin{lstlisting}[language=Python]
# Testing...
print('Testing...')
[label_test_pred,_] = model.predict(data_test)
label_test_pred = np.argmax(label_test_pred, axis = -1) + 1

# Confusion Matrix
cm = confusion_matrix(label_test, label_test_pred)
print(cm)
cm = cm.astype('float')/cm.sum(axis=1)[:, np.newaxis]
cm = np.around(cm, decimals=2)
print(cm)

# Accuracy
test_accuracy = np.sum(label_test == label_test_pred) / (label_test.shape[0])
print(test_accuracy)
\end{lstlisting}

Different from that in \sect{sec:cnn}, we load data, label, and domain in the \lstinline{load_data} function.
The domain is defined as the location of the human, which is embedded in the file name.

\begin{lstlisting}[language=Python]
def load_data(path_to_data):
    global T_MAX
    data = []
    label = []
    domain = []
    for data_root, data_dirs, data_files in os.walk(path_to_data):
        for data_file_name in data_files:

            file_path = os.path.join(data_root,data_file_name)
            try:
                data_1 = scio.loadmat(file_path)['doppler_spectrum']    # [6,121,T]
                label_1 = int(data_file_name.split('-')[1])
                location_1 = int(data_file_name.split('-')[2])
                orientation_1 = int(data_file_name.split('-')[3])
                repetition_1 = int(data_file_name.split('-')[4])
                
                # Downsample
                data_1 = data_1[:,:,0::10]

                # Select Motion
                if (label_1 not in ALL_MOTION):
                    continue

                # Normalization
                data_normed_1 = normalize_data(data_1)
                
                # Update T_MAX
                if T_MAX < np.array(data_1).shape[2]:
                    T_MAX = np.array(data_1).shape[2]                
            except Exception:
                continue

            # Save List
            data.append(data_normed_1.tolist())
            label.append(label_1)
            domain.append(location_1)
            
    # Zero-padding
    data = zero_padding(data, T_MAX)

    # Swap axes
    data = np.swapaxes(np.swapaxes(data, 1, 3), 2, 3)   # [N,6,121,T_MAX]=>[N,T_MAX,6,121]
    data = np.expand_dims(data, axis=-1)                # [N,T_MAX,6,121]=>[N,T_MAX,6,121,1]

    # Convert label and domain to ndarray
    label = np.array(label)
    domain = np.array(domain)

    # data(ndarray): [N,T_MAX,6,121,1], label(ndarray): [N,], domain(ndarray): [N,]
    return data, label, domain
\end{lstlisting}

To define the network, we use a CNN layer and an RNN layer as the feature extractor, which is similar to that in \sect{sec:rnn}.
In the gesture recognizer and domain discriminator, we use two fully-connected layers and an output layer activated by \lstinline{softmax} function, respectively.
We use categorical cross-entropy loss for both label prediction and domain prediction outputs.
The domain prediction loss is weighted with \lstinline{loss_weight_domain} and subtracted from the label prediction loss.

\begin{lstlisting}[language=Python]
def build_model(input_shape, n_class, n_domain):
    model_input = Input(shape=input_shape, dtype='float32', name='name_model_input')    # (@,T_MAX,6,121,1)

    # CNN+RNN+Adversarial
    x = TimeDistributed(Conv2D(16,kernel_size=(3,6),activation='relu',data_format='channels_last',\
        input_shape=input_shape))(model_input)              # (@,T_MAX,6,121,1)=>(@,T_MAX,4,116,16)
    x = TimeDistributed(MaxPooling2D(pool_size=(2,2)))(x)   # (@,T_MAX,4,116,16)=>(@,T_MAX,2,58,16)

    x = TimeDistributed(Flatten())(x)                       # (@,T_MAX,2,58,16)=>(@,T_MAX,2*58*16)
    x = TimeDistributed(Dense(128,activation='relu'))(x)    # (@,T_MAX,2*58*16)=>(@,T_MAX,128)
    x = TimeDistributed(Dropout(f_dropout_ratio))(x)
    x = TimeDistributed(Dense(64,activation='relu'))(x)     # (@,T_MAX,128)=>(@,T_MAX,64)

    x = GRU(n_gru_hidden_units,return_sequences=False)(x)   # (@,T_MAX,64)=>(@,64)
    x_feat = Dropout(f_dropout_ratio)(x)

    # Label prediction part
    x_1 = Dense(64, activation='relu')(x_feat)      # (@,64)=>(@,64)
    x_1 = Dense(32, activation='relu')(x_1)         # (@,64)=>(@,32)
    model_output_label = Dense(n_class, activation='softmax', name='name_model_output_label')(x_1)      # (@,32)=>(@,n_class)

    # Domain prediction part
    x_2 = Dense(64, activation='relu')(x_feat)      # (@,64)=>(@,64)
    x_2 = Dense(32, activation='relu')(x_2)         # (@,64)=>(@,32)
    model_output_domain = Dense(n_domain, activation='softmax', name='name_model_output_domain')(x_2)   # (@,32)=>(@,n_domain)


    model = Model(inputs=model_input, outputs=[model_output_label, model_output_domain])
    model.compile(optimizer=keras.optimizers.RMSprop(lr=f_learning_rate),
            loss = {'name_model_output_label':custom_loss_label(), 'name_model_output_domain':custom_loss_domain()},
            loss_weights={'name_model_output_label':1, 'name_model_output_domain':-1*loss_weight_domain},
            metrics={'name_model_output_label':'accuracy', 'name_model_output_domain':'accuracy'}
            )
    
    return model
\end{lstlisting}

The pre-defined \lstinline{custom_loss_label} and \lstinline{custom_loss_domain} are categorical crossentropy losses for both label prediction and domain prediction.

\begin{lstlisting}[language=Python]
def custom_loss_label():
    def lossfn(y_true, y_pred):
        myloss_batch = -1 * K.sum(y_true*K.log(y_pred+K.epsilon()), axis=-1, keepdims=False)
        myloss = K.mean(myloss_batch, axis=-1, keepdims=False)
        return myloss
    return lossfn
\end{lstlisting}

\begin{lstlisting}[language=Python]
def custom_loss_domain():
    def lossfn(y_true, y_pred):
        myloss_batch = -1 * K.sum(y_true*K.log(y_pred+K.epsilon()), axis=-1, keepdims=False)
        myloss = K.mean(myloss_batch, axis=-1, keepdims=False)
        return myloss
    return lossfn
\end{lstlisting}

\subsection{Complex-valued Neural Network}

In this section, we will present a more complicated wireless sensing task with deep learning.
Many wireless sensing approaches employ Fast Fourier Transform (FFT) on a time series of RF data to obtain time-frequency spectrograms of human activities. 
FFT suffers from errors due to an effect known as leakage, when the block of data is not periodic (the most common case in practice), which results in a smeared spectrum of the original signal 
and further leads to misleading data representation for learning-based sensing.
Classical approaches reduce leakage by windowing, which cannot eliminate leakage entirely.
Considering the significant fitting capability of deep neural networks, we can design a signal processing network to learn an optimal function to minimize or nearly eliminate the leakage and enhance the spectrums, which we call the Signal Enhancement Network (SEN).

\begin{figure}[tb]
\centering
\hspace{0.1in}\includegraphics[width=0.7\textwidth]{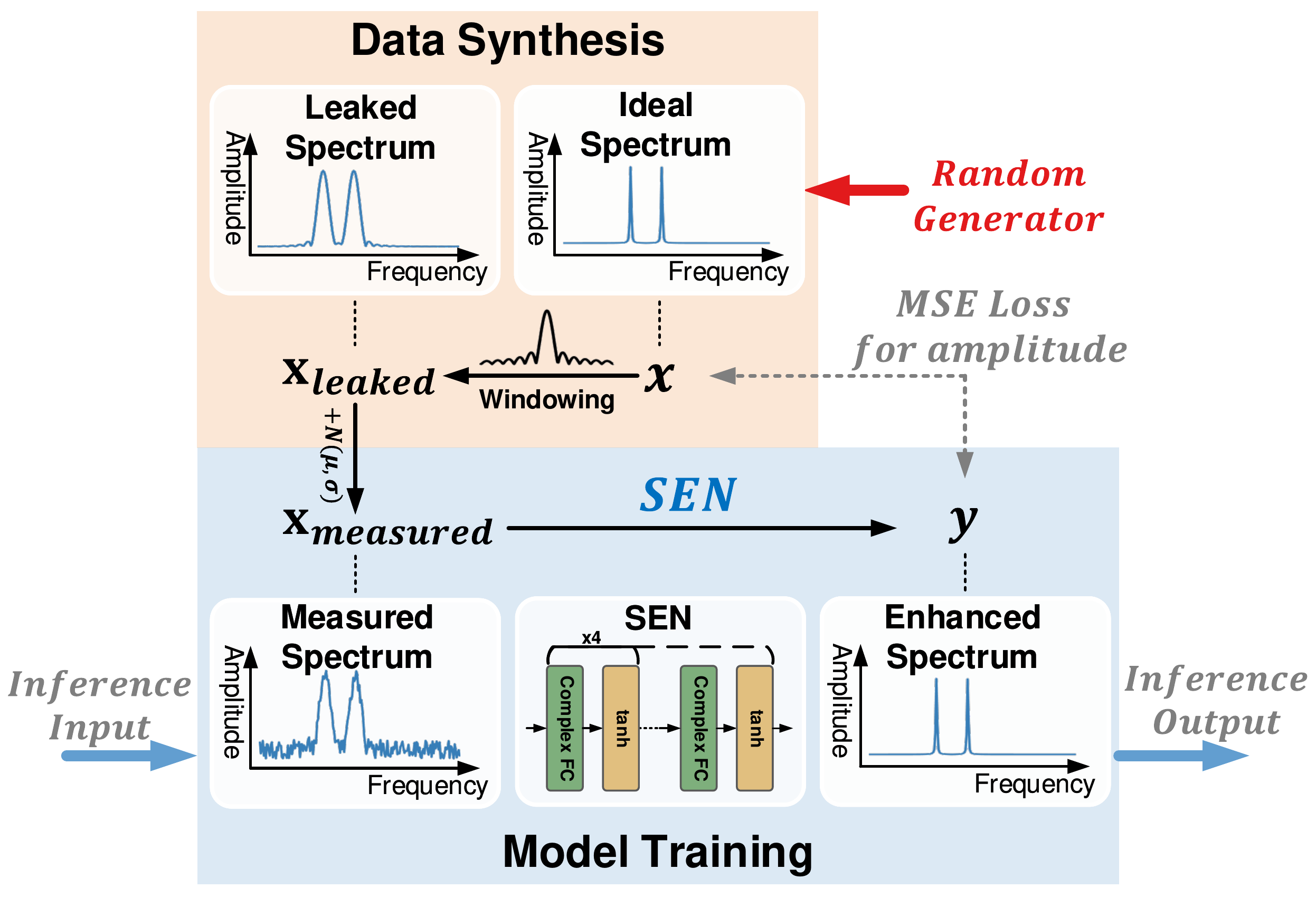}
\caption{The training process of the signal processing network.}
\label{fig:cvnn_train_proc}
\end{figure}

The signal processing network takes as input a spectrogram transformed from wireless signals via STFT, removes the spectral leakage in the spectrogram, and recovers the underlying actual frequency components.
\fig{fig:cvnn_train_proc} shows the training process of the network.
As shown in the upper part of \fig{fig:cvnn_train_proc}, we randomly generate ideal spectrums with 1 to 5 frequency components, whose amplitudes, phases, and frequencies are uniformly drawn from their ranges of interest.
Then, the ideal spectrum is converted to the leaked spectrum following the process in the following equation to simulate the windowing effect and random complex noises:
\begin{equation}
\label{eq:spec}
    \hat{\mathbf{s}} = \mathbf{A}\mathbf{s}+\mathbf{n},
\end{equation}
where $\mathbf{s}$ and $\hat{\mathbf{s}}$ are the ideal and estimated frequency spectrum, respectively, $\mathbf{n}$ represents the additive Gaussian noise vector, and $\mathbf{A}$ is the convolution matrix of the windowing function in the frequency domain.
The $i^{th}$ column of $\mathbf{A}$ is:
\begin{equation}
\label{eq:A_matrix}
    \mathbf{A_{(:,i)}} = {\rm FFT}(\mathbf{\varpi}) * \delta(i).
\end{equation}
where $\mathbf{\varpi}$ represents the windowing function of FFT in time domain.

The amplitude of the noise follows a Gaussian distribution and its phase follows a uniform distribution in $[0,2\pi]$. 
The network takes the leaked spectrum as input and outputs the enhanced spectrum close to the ideal one. 
Thus, we minimize the $L_2$ loss $L=||{\rm SEN}(\hat{\mathbf{s}})-\mathbf{s}||_2$ during training.
During inference, the spectrums measured from real-world scenarios are
normalized to $[0,1]$ and fed into the network to obtain the enhanced spectrum.

We now present the implementation code in detail.

\begin{figure}[!t]
\centering
    \subfloat[][\label{subfig:real_neuron}]{
      \includegraphics[width=0.5\textwidth]{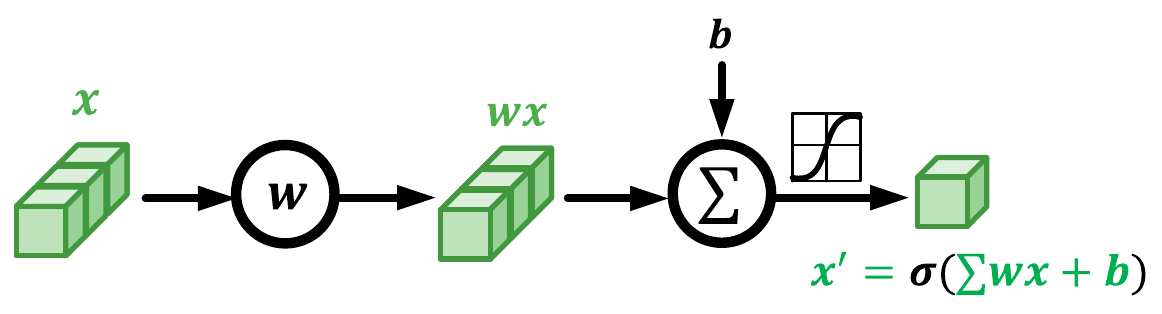}
    }\\
    \vspace{-0.15in}
    \subfloat[][\label{subfig:complex_neuron}]{
      \includegraphics[width=0.5\textwidth]{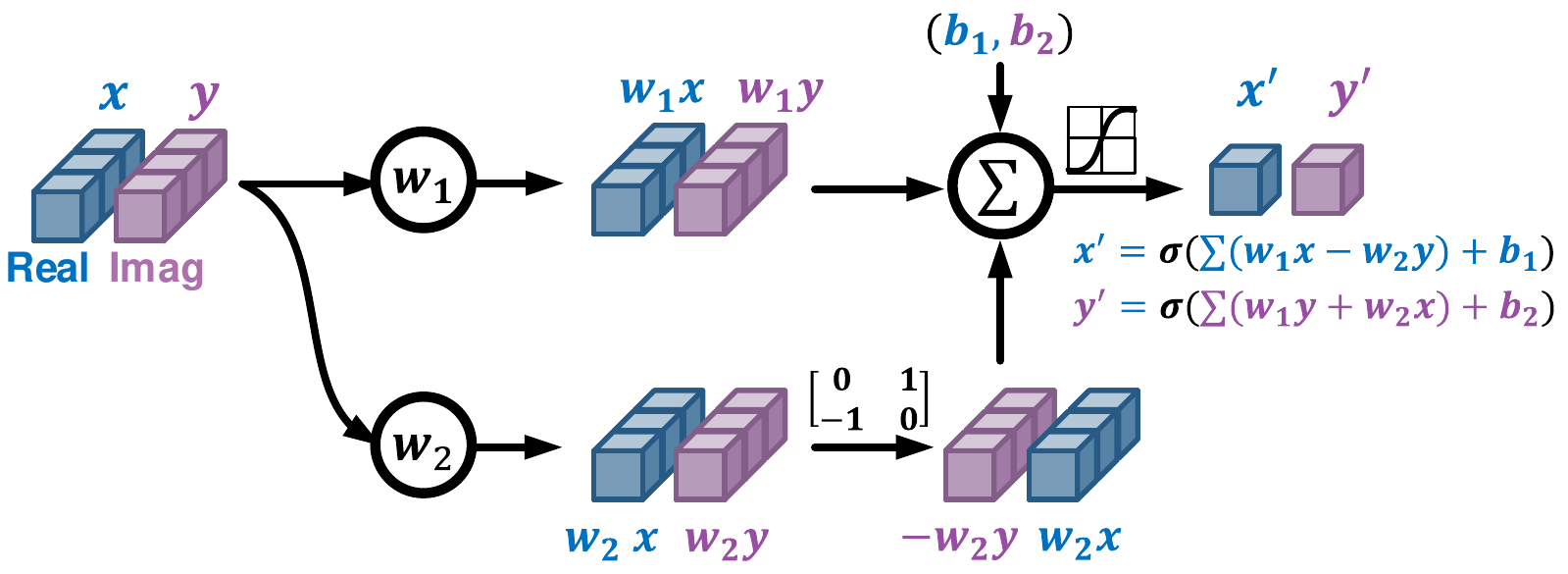}
    }
    \vspace{-0.15in}
    \caption{Comparison between (a) real-valued and (b) complex-valued neurons.}
    \label{fig:neuron}
\end{figure}

\begin{figure}[tb]
\centering
\hspace{0.1in}\includegraphics[width=0.49\textwidth]{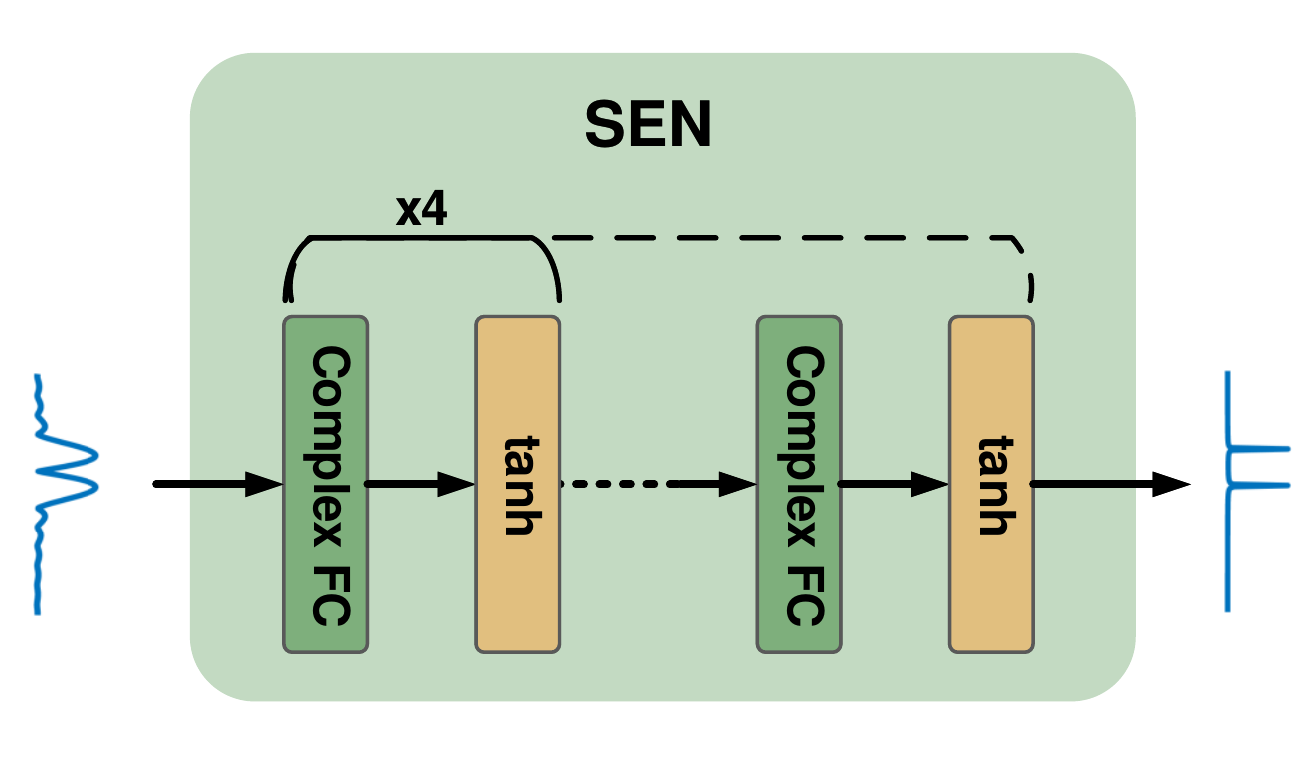}
\caption{Complex-valued neural network architecture.}
\label{fig:cvnn_arch}
\end{figure}

Different from previous sections, we use the PyTorch platform to implement the network.
PyTorch provides the interface to implement custom layers, which makes the implementation of the CVNN much more convenient.
We first import some necessary packages as follows.

\begin{lstlisting}[language=Python]
import os,sys,math,scipy,imp
import numpy as np
import scipy.io as scio
import torch, torchvision
import torch.nn as nn
from scipy import signal
from math import sqrt,log,pi
from torch.fft import fft,ifft
from torch.nn.functional import relu, softmax, cross_entropy
from torch import sigmoid,tanh
from torch.nn import MSELoss as MSE
\end{lstlisting}

Some parameters are defined in this part.

\begin{lstlisting}[language=Python]
# Definition
wind_len = 125
wind_type = 'gaussian'
n_max_freq_component = 3
AWGN_amp = 0.01
str_modelname_prefix = './SEN_Results/SEN_' + wind_type + '_W' + str(wind_len)
str_model_name_pretrained = str_modelname_prefix + '_E1000.pt'
feature_len = 121
padded_len = 1000
crop_len = feature_len
blur_matrix_left = []

# Hyperparameters
n_begin_epoch = 1
n_epoch = 10000
n_itr_per_epoch = 500
n_batch_size = 64
n_test_size = 200
f_learning_rate = 0.001
\end{lstlisting}

The program begins with the following code.
We first generate the convolution matrix of the windowing function (\eq{eq:A_matrix}) with the pre-defined function \lstinline{generate_blur_matrix_complex}, which will be introduced shortly.
Then we define the SEN model and move it to the GPU processor with the API function \lstinline{cuda}.
After the model definition, we train the model with synthetic spectrograms, during which process we save the trained model every 500 epochs.
The trained and saved model can be directly loaded and used to enhance spectrograms.

\begin{lstlisting}[language=Python]
if __name__ == "__main__":
    # Generate blur matrix
    blur_matrix_right = generate_blur_matrix_complex(wind_type=wind_type, wind_len=wind_len, padded_len=padded_len, crop_len=crop_len)

    # Define model
    print('Model building...')
    model = SEN(feature_len=feature_len)
    model.cuda()

    # Train model
    print('Model training...')
    train(model=model, blur_matrix_right=blur_matrix_right, feature_len=feature_len, n_epoch=n_epoch, n_itr_per_epoch=n_itr_per_epoch, n_batch_size=n_batch_size, optimizer=torch.optim.RMSprop(model.parameters(), lr=f_learning_rate))

\end{lstlisting}

This \lstinline{generate_blur_matrix_complex} function is used to generate the convolution matrix of the windowing function.
The core idea behind this function is to enumerate all the frequencies, apply the window function on the sinusoid signal, and generate the corresponding spectrums with FFT.
After obtaining the convolution matrix, we can bridge the gap between the ideal and the leaked spectrograms with \eq{eq:spec}.
In other words, we can directly get the leaked spectrograms by multiplying the idea spectrograms with the convolution matrix.

\begin{lstlisting}[language=Python]
def generate_blur_matrix_complex(wind_type, wind_len=251, padded_len=1000, crop_len=121):
    # Generate matrix used to introduce spec leakage in complex domain
    # ret: (ndarray.complex128) [crop_len, crop_len](row first)
    # Row first: each row represents the spectrum of one single carrier

    # Steps: carrier/windowing/pad/fft/crop/unwrap/norm

    # Parameters offloading
    fs = 1000
    n_f_bins = crop_len
    f_high = int(n_f_bins/2)
    f_low = -1 * f_high
    init_phase = 0

    # Carrier
    t_ = np.arange(0,wind_len).reshape(1,wind_len)/fs       # [1,wind_len] (0~wind_len/fs seconds)
    freq = np.arange(f_low,f_high+1,1).reshape(n_f_bins,1)  # [n_f_bins,1] (f_low~f_high Hz)
    phase = 2 * pi * freq * t_ + init_phase                 # [n_f_bins,wind_len]
    signal = np.exp(1j*phase)                               # [n_f_bins,wind_len]~[121,251]

    # Windowing
    if wind_type == 'gaussian':
        window = scipy.signal.windows.gaussian(wind_len, (wind_len-1)/sqrt(8*log(200)), sym=True)   # [wind_len,]
    else:
        window = scipy.signal.get_window(wind_type, wind_len)
    sig_wind = signal * window       # [n_f_bins,wind_len]*[wind_len,]=[n_f_bins,wind_len]~[121,251]

    # Pad/FFT
    sig_wind_pad = np.concatenate((sig_wind, np.zeros((n_f_bins,padded_len-wind_len))),axis=1)  # [n_f_bins,wind_len]=>[n_f_bins,padded_len]
    sig_wind_pad_fft = np.fft.fft(sig_wind_pad, axis=-1)    # [n_f_bins,padded_len]~[121,1000]

    # Crop
    n_freq_pos = f_high + 1
    n_freq_neg = abs(f_low)
    sig_wind_pad_fft_crop = np.concatenate((sig_wind_pad_fft[:,:n_freq_pos],\
        sig_wind_pad_fft[:,-1*n_freq_neg:]), axis=1)      # [n_f_bins,crop_len]~[121,121]

    # Unwrap
    n_shift = n_freq_neg
    sig_wind_pad_fft_crop_unwrap = np.roll(sig_wind_pad_fft_crop, shift=n_shift, axis=1) # [n_f_bins,crop_len]~[121,121]

    # Norm (amp_max=1)
    _sig_amp = np.abs(sig_wind_pad_fft_crop_unwrap)
    _sig_ang = np.angle(sig_wind_pad_fft_crop_unwrap)
    _max = np.tile(_sig_amp.max(axis=1,keepdims=True), (1,crop_len))
    _min = np.tile(_sig_amp.min(axis=1,keepdims=True), (1,crop_len))
    _sig_amp_norm = _sig_amp / _max
    sig_wind_pad_fft_crop_unwrap_norm = _sig_amp_norm * np.exp(1j*_sig_ang)

    # Return
    ret = sig_wind_pad_fft_crop_unwrap_norm

    return ret
\end{lstlisting}

This function is to generate one batch of spectrograms in both the leaked form and the idea form.
The process is the same as that illustrated in \eq{eq:spec}.

\begin{lstlisting}[language=Python]
def syn_one_batch_complex(blur_matrix_right, feature_len, n_batch):
    # Syn. HiFi, blurred and AWGN signal in complex domain
    # ret: (ndarray.complex128) [@,feature_len]
    # blur_matrix_right: Row first (each row represents the spectrum of one single carrier)

    # Syn. x [@,feature_len]
    x = np.zeros((n_batch, feature_len))*np.exp(1j*0)
    for i in range(n_batch):
        num_carrier = int(np.random.randint(0,n_max_freq_component,1))
        idx_carrier = np.random.permutation(feature_len)[:num_carrier]
        x[i,idx_carrier] = np.random.rand(1,num_carrier) * np.exp(1j*( 2*pi*np.random.rand(1,num_carrier) - pi ))

    # Syn. x_blur [@,feature_len]
    x_blur = x @ blur_matrix_right

    # Syn. x_tilde [@,feature_len]
    x_tilde = x_blur + 2*AWGN_amp*(np.random.random(x_blur.shape)-0.5) *\
        np.exp(1j*( 2*pi*np.random.random(x_blur.shape) - pi ))

    return x, x_blur, x_tilde
\end{lstlisting}

This part demonstrates the code on how to implement the SEN network.
According to the API of PyTorch, both the \lstinline{__init__} and the \lstinline{forward} interfaces should be implemented with customized algorithms.
In the \lstinline{__init__} function, we defined five complex-valued fully-connected layers.
In the \lstinline{forward} function, we defined the network structure by concatenating the five FC layers and specifying the input and output layers.

\begin{lstlisting}[language=Python]
class SEN(nn.Module):
    def __init__(self, feature_len):
        super(SEN, self).__init__()
        self.feature_len = feature_len
        
        self.fc_1 = m_Linear(feature_len, feature_len)
        self.fc_2 = m_Linear(feature_len, feature_len)
        self.fc_3 = m_Linear(feature_len, feature_len)
        self.fc_4 = m_Linear(feature_len, feature_len)
        self.fc_out = m_Linear(feature_len, feature_len)

    def forward(self, x):
        h = x   # (@,*,2,H)

        h = tanh(self.fc_1(h))          # (@,*,2,H)=>(@,*,2,H)
        h = tanh(self.fc_2(h))          # (@,*,2,H)=>(@,*,2,H)
        h = tanh(self.fc_3(h))          # (@,*,2,H)=>(@,*,2,H)
        h = tanh(self.fc_4(h))          # (@,*,2,H)=>(@,*,2,H)
        output = tanh(self.fc_out(h))   # (@,*,2,H)=>(@,*,2,H)

        return output
\end{lstlisting}

This \lstinline{m_Linear} class leverages the interface of PyTorch to define the customized complex-valued fully-connected layer, which is the implementation of the network structure illustrated in \fig{fig:neuron}.

\begin{lstlisting}[language=Python]
class m_Linear(nn.Module):
    def __init__(self, size_in, size_out):
        super().__init__()
        self.size_in, self.size_out = size_in, size_out

        # Creation
        self.weights_real = nn.Parameter(torch.randn(size_in, size_out, dtype=torch.float32))
        self.weights_imag = nn.Parameter(torch.randn(size_in, size_out, dtype=torch.float32))
        self.bias = nn.Parameter(torch.randn(2, size_out, dtype=torch.float32))

        # Initialization
        nn.init.xavier_uniform_(self.weights_real, gain=1)
        nn.init.xavier_uniform_(self.weights_imag, gain=1)
        nn.init.zeros_(self.bias)
    
    def swap_real_imag(self, x):
        # [@,*,2,Hout]
        # [real, imag] => [-1*imag, real]
        h = x                   # [@,*,2,Hout]
        h = h.flip(dims=[-2])   # [@,*,2,Hout]  [real, imag]=>[imag, real]
        h = h.transpose(-2,-1)  # [@,*,Hout,2]
        h = h * torch.tensor([-1,1]).cuda()     # [@,*,Hout,2] [imag, real]=>[-1*imag, real]
        h = h.transpose(-2,-1)  # [@,*,2,Hout]
        
        return h

    def forward(self, x):
        # x: [@,*,2,Hin]
        h = x           # [@,*,2,Hin]
        h1 = torch.matmul(h, self.weights_real) # [@,*,2,Hout]
        h2 = torch.matmul(h, self.weights_imag) # [@,*,2,Hout]
        h2 = self.swap_real_imag(h2)            # [@,*,2,Hout]
        h = h1 + h2                             # [@,*,2,Hout]
        h = torch.add(h, self.bias)             # [@,*,2,Hout]+[2,Hout]=>[@,*,2,Hout]
        return h
\end{lstlisting}

This \lstinline{loss_function} defines the loss function of the SEN network.
The loss is defined as the Euclidean distance between the idea spectrum and the netwrok predicted spectrum.
Only the amplitude of the spectrums are considered.

\begin{lstlisting}[language=Python]
def loss_function(x, y):
    # x,y: [@,*,2,H]
    x = torch.linalg.norm(x,dim=-2) # [@,*,2,H]=>[@,*,H]
    y = torch.linalg.norm(y,dim=-2) # [@,*,2,H]=>[@,*,H]

    # MSE loss for Amp
    loss_recon = MSE(reduction='mean')(x, y)
    return loss_recon
\end{lstlisting}

In this \lstinline{train} function, we implement the training process of SEN as that described in \fig{fig:cvnn_train_proc}.
In each epoch, we generate and train the network with multiple iterations.
For each iteration, we generate a batch of synthetic spectrums in leaked format.
Each leaked spectrum have an idea spectrum as the label.

\begin{lstlisting}[language=Python]
def train(model, blur_matrix_right, feature_len, n_epoch, n_itr_per_epoch, n_batch_size, optimizer):
    for i_epoch in range(n_begin_epoch, n_epoch+1):
        model.train()
        total_loss_this_epoch = 0
        for i_itr in range(n_itr_per_epoch):
            x, _, x_tilde = syn_one_batch_complex(blur_matrix_right=blur_matrix_right, feature_len=feature_len, n_batch=n_batch_size)
            x = complex_array_to_bichannel_float_tensor(x)
            x_tilde = complex_array_to_bichannel_float_tensor(x_tilde)
            x = x.cuda()
            x_tilde = x_tilde.cuda()

            optimizer.zero_grad()
            y = model(x_tilde)
            loss = loss_function(x, y)
            loss.backward()
            optimizer.step()
            
            total_loss_this_epoch += loss.item()
            
            if i_itr % 10 == 0:
                print('--------> Epoch: {}/{} loss: {:.4f} [itr: {}/{}]'.format(
                    i_epoch+1, n_epoch, loss.item() / n_batch_size, i_itr+1, n_itr_per_epoch), end='\r')
        
        # Validate
        model.eval()
        x, _, x_tilde = syn_one_batch_complex(blur_matrix_right=blur_matrix_right, feature_len=feature_len, n_batch=n_batch_size)
        x = complex_array_to_bichannel_float_tensor(x)
        x_tilde = complex_array_to_bichannel_float_tensor(x_tilde)
        x = x.cuda()
        x_tilde = x_tilde.cuda()
        y = model(x_tilde)
        total_valid_loss = loss_function(x, y)
        print('========> Epoch: {}/{} Loss: {:.4f}'.format(i_epoch+1, n_epoch, total_valid_loss) + ' ' + wind_type + '_' + str(wind_len) + ' '*20)

        if i_epoch % 500 == 0:
            torch.save(model, str_modelname_prefix+'_E'+str(i_epoch)+'.pt')
\end{lstlisting}

This function converts the complex-valued arrays to double channel real-valued arrays.
This is because the GPU only supports the calculations of real numbers.
We use a little trick to implement the complex-valued network by separating the real and imaginary parts into two real arrays.

\begin{lstlisting}[language=Python]
def complex_array_to_bichannel_float_tensor(x):
    # x: (ndarray.complex128) [@,*,H]
    # ret: (tensor.float32) [@,*,2,H]
    x = x.astype('complex64')
    x_real = x.real     # [@,*,H]
    x_imag = x.imag     # [@,*,H]
    ret = np.stack((x_real,x_imag), axis=-2)    # [@,*,H]=>[@,*,2,H]
    ret = torch.tensor(ret)
    return ret
\end{lstlisting}

This function converts the double channel real-valued arrays into complex-valued arrays.

\begin{lstlisting}[language=Python]
def bichannel_float_tensor_to_complex_array(x):
    # x: (tensor.float32) [@,*,2,H]
    # ret: (ndarray.complex64) [@,*,H]
    x = x.numpy()
    x = np.moveaxis(x,-2,0)  # [@,*,2,H]=>[2,@,*,H]
    x_real = x[0,:]
    x_imag = x[1,:]
    ret = x_real + 1j*x_imag
    return ret
\end{lstlisting}

After training the SEN network with sufficient epochs, we test the performance with spectrograms collected from Wi-Fi.

First, we define some parameters.
The STFT window width is set to 125, and the window type is set to ``gaussian''.
The path to the pre-trained model and the CSI file is selected.

\begin{lstlisting}[language=Python]
W = 125
wind_type = 'gaussian'
str_model_name = './SEN_Results/SEN_' + wind_type + '_W' + str(W) + '_E500.pt'
file_path_csi = 'Widar3_data/20181130/user1/user1-1-1-1-1.mat'
\end{lstlisting}

The program begins with loading the pre-trained model.
Then, the CSI data is loaded and transformed to spectrograms with the predefined function \lstinline{csi_to_spec}.
After that, the complex-valued spectrograms are transformed to double real-valued channel tensors and processed with the SEN model.
The results and the raw spectrograms are stored in files.

\begin{lstlisting}[language=Python]
if __name__ == "__main__":
    # Load trained model
    print('Loading model...')
    model = torch.load(str_model_name)

    print('Testing model...')
    model.eval()
    with torch.no_grad():
        # Import raw spectrogram
        data_1 = csi_to_spec()
        
        # Enhance spectrogram
        x_tilde = complex_array_to_bichannel_float_tensor(data_1)   # [6,121,T]=>[6,121,2,T]
        x_tilde = x_tilde.permute(0,3,2,1)              # [6,121,2,T]=>[6,T,2,121]
        y = model(x_tilde.cuda()).cpu()                 # [6,T,2,121]
        y = bichannel_float_tensor_to_complex_array(y)  # [6,T,121]
        y = np.transpose(y,(0,2,1))                     # [6,T,121]=>[6,121,T]
        scio.savemat('SEN_test_x_tilde_complex_W' + str(W) + '.mat', {'x_tilde':data_1})
        scio.savemat('SEN_test_y_complex_W' + str(W) + '.mat', {'y':y})
\end{lstlisting}

This function first transforms the CSI data into spectrograms and crops the concerned frequency range between $[-60, 60]$ Hz.
Then, it unwraps the frequency bins and performs normalizations.

\begin{lstlisting}[language=Python]
def csi_to_spec():
    global file_path_csi
    global W
    signal = scio.loadmat(file_path_csi)['csi_mat'].transpose() # [6,T] complex
    # STFT
    _, spec = STFT(signal, fs=1000, stride=1, wind_wid=W, dft_wid=1000, window_type='gaussian') # [6,1000,T]j
    # Crop
    spec_crop = np.concatenate((spec[:,:61], spec[:,-60:]), axis=1) # [1,1000,T]j=>[1,121,T]j
    # Unwrap
    spec_crop_unwrap = np.roll(spec_crop, shift=60, axis=1) # [1,121,T]j
    # Normalize
    spec_crop_unwrap_norm = normalize_data(spec_crop_unwrap)     # [6,121,T] complex
    if np.sum(np.isnan(spec_crop_unwrap_norm)):
        print('>>>>>>>>> NaN detected!')
    ret = spec_crop_unwrap_norm
    return ret
\end{lstlisting}

This function transforms the time domain CSI data into frequency domain spectrograms with the API function \lstinline{scipy.signal.stft}.

\begin{lstlisting}[language=Python]
def STFT(signal, fs=1, stride=1, wind_wid=5, dft_wid=5, window_type='gaussian'):
    assert dft_wid >= wind_wid and wind_wid > 0 and stride <= wind_wid and stride > 0\
        and isinstance(stride, int) and isinstance(wind_wid, int) and isinstance(dft_wid, int)\
        and isinstance(fs, int) and fs > 0

    if window_type == 'gaussian':
        window = scipy.signal.windows.gaussian(wind_wid, (wind_wid-1)/sqrt(8*log(200)), sym=True)
    elif window_type == 'rect':
        window = np.ones((wind_wid,))
    else:
        window = scipy.signal.get_window(window_type, wind_wid)
    
    f_bins, t_bins, stft_spectrum = scipy.signal.stft(x=signal, fs=fs, window=window, nperseg=wind_wid, noverlap=wind_wid-stride, nfft=dft_wid,\
        axis=-1, detrend=False, return_onesided=False, boundary='zeros', padded=True)
    
    return f_bins, stft_spectrum
\end{lstlisting}

This function scales the spectrograms to normalize the values into $[0, 1]$.

\begin{lstlisting}[language=Python]
def normalize_data(data_1):
    # max=1
    # data(ndarray.complex)=>data_norm(ndarray.complex): [6,121,T]=>[6,121,T]
    data_1_abs = abs(data_1)
    data_1_max = data_1_abs.max(axis=(1,2),keepdims=True)     # [6,121,T]=>[6,1,1]
    data_1_max_rep = np.tile(data_1_max,(1,data_1_abs.shape[1],data_1_abs.shape[2]))    # [6,1,1]=>[6,121,T]
    data_1_norm = data_1 / data_1_max_rep
    return  data_1_norm
\end{lstlisting}

\begin{figure}[t]
    \centering
    \subfloat[][\label{subfig:spec_ges_1}]{
      \includegraphics[width=0.4\textwidth]{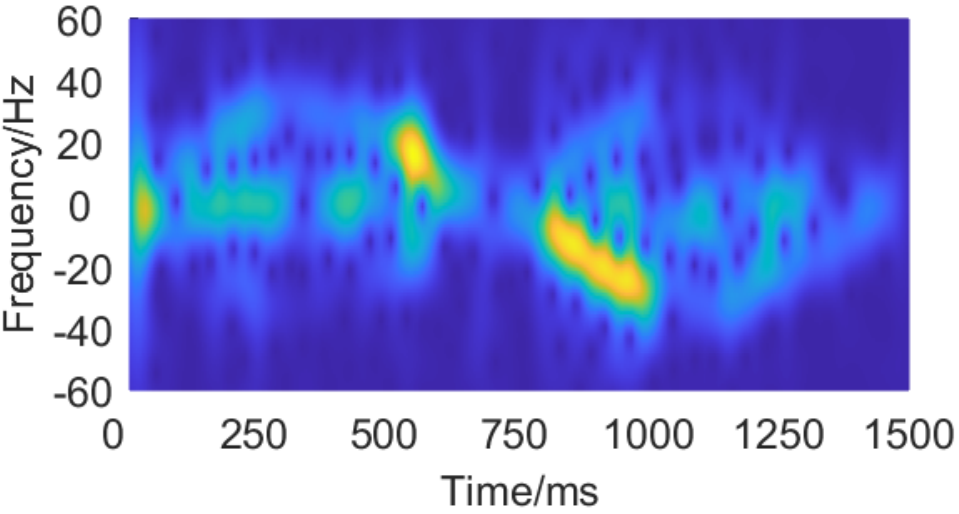}
    }
    \subfloat[][\label{subfig:spec_ges_2}]{
      \includegraphics[width=0.4\textwidth]{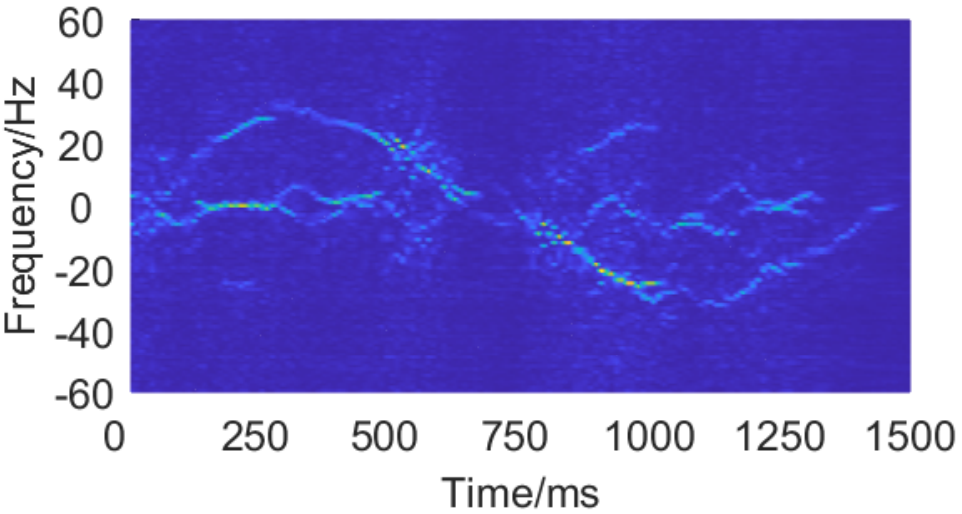}
    }
    \vspace{-0.15in}
    \caption{Illustration of the spectrogram of a pushing and pulling gesture. (a) The measured spectrogram and (b) the enhanced spectrogram from the SEN.}
    \label{fig:spec_gesture}
\end{figure}

\fig{fig:spec_gesture} demonstrates the raw and enhanced spectrograms of pushing and pulling gestures.

\bibliographystyle{ACM-Reference-Format.bst}
\bibliography{Bib/references}

\end{document}